\documentclass[runningheads]{llncs}
\usepackage[T1]{fontenc}
\usepackage{graphicx}
\usepackage{cite}
\usepackage{amssymb,amsfonts}
\usepackage{algorithm}   
\usepackage{algorithmicx} 
\usepackage[dvipsnames]{xcolor} 
\usepackage{algpseudocode}
\usepackage{courier} 

\usepackage{enumitem}
\definecolor{M_green}{HTML}{00A942}

\usepackage{graphicx}
\usepackage{textcomp}
\usepackage{xcolor}
\def\BibTeX{{\rm B\kern-.05em{\sc i\kern-.025em b}\kern-.08em
    T\kern-.1667em\lower.7ex\hbox{E}\kern-.125emX}}

\usepackage{multirow}
\usepackage{lineno,hyperref}
\modulolinenumbers[5]
\usepackage{array}
\newcolumntype{P}[1]{>{\centering\arraybackslash}p{#1}}

\begin{document}
\title{A Hybrid Heuristic Framework for Resource-Efficient Querying of Scientific Experiments Data}

%
%
\author{Mayank Patel\inst{2,1}\orcidID{0000-0002-7804-4017} \and
Minal Bhise\inst{1}\orcidID{0000-0003-4364-3930} }
\authorrunning{M. Patel et al.}
%
\institute{Distributed Databases Research Group, Dhirubhai Ambani University 
\email{mayank,minal\_bhise@dau.ac.in}\\
\url{https://sites.google.com/view/ddrg-dau/} \and {Adani University}
}

\maketitle              
\begin{abstract}
Scientific experiments and modern applications are generating large amounts of data every day. Most organizations utilize In-house servers or Cloud resources to manage application data and workload. 
The traditional database management system (DBMS) and  HTAP systems spend significant time \& resources to load the entire dataset into DBMS before starting query execution. On the other hand, in-situ engines may reparse required data multiple times, increasing resource utilization and data processing costs.
Additionally, over or under allocation of resources also increases application running costs. 
    This paper proposes a lightweight Resource Availability \& Workload aware Hybrid Framework (RAW-HF) to optimize querying raw data by utilizing existing finite resources efficiently.
RAW-HF includes modules that help optimize the resources required to execute a given workload and maximize the utilization of existing resources. 
The impact of applying RAW-HF to real-world scientific dataset workloads like Sloan Digital Sky Survey (SDSS) and Linked Observation Data (LOD) presented over 90\% and 85\% reduction in workload execution time (WET) compared to widely used traditional DBMS PostgreSQL. 
The overall CPU, IO resource utilization, and WET have been reduced by 26\%, 25\%, and 26\%, respectively, while improving memory utilization by 33\%, compared to the state-of-the-art workload aware partial loading technique (WA) proposed for hybrid systems. A comparison of MUAR technique used by RAW-HF with machine learning based resource allocation techniques like PCC is also presented.

\keywords{Big Data Partitioning \and Optimize Resource Utilization \and Raw Data Query Processing \and Real-time Dynamic Resource Allocation \and Task Scheduling.}
\end{abstract}
%
%
%
\section{Introduction}

The data generation speed of modern applications, scientific experiments, and IoT applications is increasing rapidly. 
The volume of Astronomy datasets like the Sloan Digital Sky Survey (SDSS) has increased by 233 times when comparing the first version (DR-1) released in 2003 to the most recent version (DR-17) released in 2021  \cite{Abdurro_uf_Accetta_Aerts_Silva_Aguirre_Ahumada_Ajgaonkar_Filiz_Ak_Alam_Allende_Prieto_Almeida_et_al_2022}. 
NASA’s Earth Observing System (EOS) collects over 3.3TB of data daily from more than 30 polar-orbiting and low inclination satellites   \cite{Guo_Wang_Liang_2016}.
Traditional database management systems require loading the entire dataset into DBMS before executing a single query, requiring a significant amount of time and resources upfront. 

A research work concluded that slow IO devices like magnetic disks are the primary bottleneck in most DBMS data loading operations \cite{Dziedzic_Karpathiotakis_Alagiannis_Appuswamy_Ailamaki_2017}. 
Therefore, most traditional and modern database management systems cannot utilize the existing CPU resources. A study observed that only 12\% of CPUs are utilized at data centers \cite{Ailamaki_2015,Maximilien_Hadas_DanducciII_Moser_2022}. 
This underutilization of resources increases application running costs for cloud and in-house distributed environments. A. Dziedzic  et al.\cite{Dziedzic_Karpathiotakis_Alagiannis_Appuswamy_Ailamaki_2017} and M. Patel  et al.\cite{Patel_Bhise_2023a} have observed that loading data in parallel cannot reduce the data loading time for systems with disk-based storage devices. Therefore, researchers developed in-situ engines with main memory caching and indexing features to query raw data directly, eliminating any need to load the data \cite{Alagiannis_Borovica_Branco_Idreos_Ailamaki_2012,Olma_Karpathiotakis_Alagiannis_Athanassoulis_Ailamaki_2020}. However, query execution time (QET) of initial queries is much high for in-situ engines as they have to process the raw data after the arrival of a query. 

A framework is required to manage the data loading, query execution, resource allocation, and task scheduling operations that do not utilize resources unnecessarily. It should also utilize resources effectively to avoid underutilization as well. ARMFUL queries multiple raw files in parallel to reduce data to result time and improve resource utilization \cite{Silva_Leite_Camata_deOliveira_Coutinho_Valduriez_Mattoso_2017}. SCANRAW and QCA work proposed using hybrid systems consisting of DBMS \& in-situ engine that can load data in parallel to query execution on raw data to improve resource utilization  \cite{Cheng_Rusu_2015,Patel_Bhise_2022}.  
SCANRAW monitors real-time CPU and IO hardware utilization to find idle time and schedules data loading tasks, reducing repetitive processing of raw data issue of in-situ engines. 

Researchers have started considering available hardware resources or past resource utilization information during query planning to improve QET or reduce resource utilization costs for cloud \cite{Li_Wang_Wang_Sun_Peng_2022,Pimpley_Li_Sen_Srinivasan_Jindal_2022,Viswanathan_Jindal_Karanasos_2018,MUAR}. Knowledge of existing hardware resources and resources required to execute a given data loading or query task can help find resource-efficient solutions \cite{Pimpley_Li_Sen_Srinivasan_Jindal_2022}. 
Researchers and organizations have been using hybrid systems to reduce data to query time, improve QET, and efficiently utilize existing resources \cite{Abouzied_Abadi_Silberschatz_2013,Cheng_Rusu_2015,Raza_Chrysogelos_Anadiotis_Ailamaki_2020,Vyas_Panchal_Patel_Bhise_2019}. HTAP or hybrid systems utilize additional resources in processing the same dataset twice. Widely used DBMSs like Postgres, MySQL, Oracle, AutoSteer \cite{anneser2023autosteer} and other open-source systems or frameworks do not monitor the utilization of available resources or consider them during query planning. 
The existing systems or techniques proposed to address resource optimization or maximization issues have been developed for specific DBMSs, which may not work for most DBMSs, hybrid systems, or cloud vendor \cite{Pimpley_Li_Sen_Srinivasan_Jindal_2022, Kaviani_Kalinin_Maximilien_2019}. 
Therefore, this paper proposes to develop a complete framework that can find optimal ways of processing a given dataset while utilizing existing resources effectively for most DBMS or hybrid systems. 

\subsection{Motivation}
Majority of existing DBMSs and current cloud resource utilization strategies cannot utilize all available resources effectively \cite{Ailamaki_2015, Maximilien_Hadas_DanducciII_Moser_2022}. Any modern data processing system needs to handle tasks like identifying if tasks can be executed in parallel, partitioning tasks into sub-tasks, and merging them to produce final results. 
Many existing systems like Hadoop, Cloudera, PCC \cite{Pimpley_Li_Sen_Srinivasan_Jindal_2022}, PDC \cite{Tang_Byna_Dong_Koziol_2020}, and other modern systems can utilize modern hardware efficiently. However, they are heavy and consume significant resources after tasks like data-query partitioning and allocating query-specific resources.   
Optimizing hardware resource utilization is gaining attention as many cloud or hardware resources are required to query large scientific, IoT, and other modern application datasets to reduce costs. External machine learning(ML) based solutions also require significant offline time to analyze resource utilization data or train ML models to find resource-efficient ways to process large datasets \cite{Ailamaki_2015,Pimpley_Li_Sen_Srinivasan_Jindal_2022}. Most systems do not consider the real-time availability of existing resources to schedule tasks and allocate resources dynamically during run time. 

 \begin{figure}[htbp]
\centerline{\includegraphics[width=0.5\textwidth]{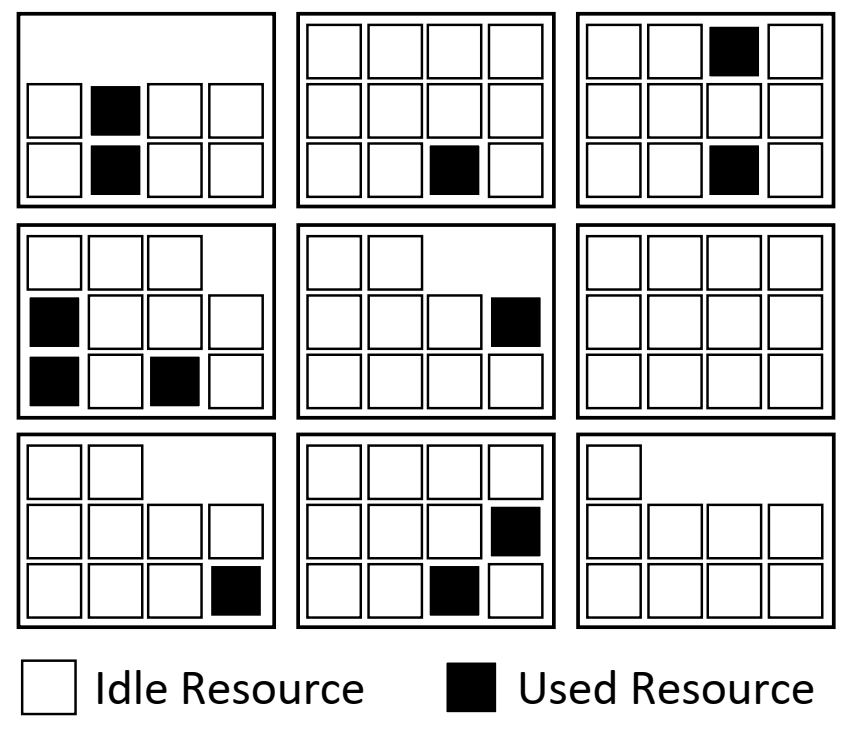}}
\caption{Resource Utilization in Cloud \cite{Maximilien_Hadas_DanducciII_Moser_2022}}
\label{Figure1_Idle_Resources}
\end{figure}

\subsection{Paper Contributions}

\begin{itemize}
\item This work proposes a Resource Availability and Workload aware Hybrid
Framework (RAW-HF) to query raw data efficiently. RAW-HF consists of query processing and resource monitoring module, optimization module ORR, and maximization module MUAR.

\item RAW-HF is able to query raw data directly saving on loading time. 

\item Using query complexity and resource utilization information RAW-HF is able to improve query execution time for simple as well as complex queries.  

\item Developed workload aware optimization module oRR which considers query complexity \& storage budget.

\item
Developed a maximization module MUAR, which does lightweight task scheduling and resource allocation based on available resources. 

 \item
Demonstrated the robustness of RAW-HF framework by optimizing resource utilization for opposite extreme real-world datasets like Sloan Digital Sky Survey (SDSS) a broad dataset and narrow ones like Linked Observation Data (LOD).  


\item Qualitatively \& quantitatively comparison of RAW-HF with state-of-the-art in-situ engine\cite{Alagiannis_Borovica_Branco_Idreos_Ailamaki_2012}, row store DBMS PostgreSQL\cite{PostgreSQL}, and workload-aware Partial Loading technique\cite{Zhao_Cheng_Rusu_2015b} is presented.

\end{itemize}
\section{Related Work}

Traditional database management systems require loading the entire dataset before executing a query. In comparison, in-situ engines can produce results faster by eliminating data loading time(DLT) and processing only required raw data after the arrival of a query. However, query execution time(QET) is much slower in in-situ engines compared to DBMSs due to reparsing of raw data. NoDB proposed to cache and index the processed raw data in the main memory to improve QET \cite{Alagiannis_Borovica_Branco_Idreos_Ailamaki_2012}. Like NoDB, SCOPE \cite{Chaiken_Jenkins_Larson_Ramsey_Shakib_Weaver_Zhou_2008}, RAW\cite{Karpathiotakis_Branco_Alagiannis_Ailamaki_2014}, Proteus \cite{Karpathiotakis_Alagiannis_Ailamaki_2016}, DaskDB \cite{DaskDB}, MySQL CSV engines, and Oracle External Table \cite{Witkowski_Colgan_Brumm_Cruanes_Baer_2011} feature help in executing SQL queries directly on CSV or JSON data files. Proteus is capable of executing queries on CSV, JSON, and relational files providing a single interface to users\cite{Karpathiotakis_Alagiannis_Ailamaki_2016}.

Slalom introduced logical partitioning and partition specific indexing to improve QET and reduce main memory utilization \cite{Olma_Karpathiotakis_Alagiannis_Athanassoulis_Ailamaki_2017,Olma_Karpathiotakis_Alagiannis_Athanassoulis_Ailamaki_2020}. LBSD \cite{AmiIdeas2021} proposes a semantic-aware, load-balanced RDF graph partitioning with partial replication, achieving up to 71\% query execution time gain over existing techniques. Data Vaults cached column data into array structures to reduce analytical query processing time \cite{Ivanova_Kersten_Manegold_2012}. However, caching entire columns consumes a large amount of main memory space. 
Therefore, S. Palkar et al. proposed to cache only one column completely and filter the remaining column rows based on the query condition to reduce memory consumption \cite{Palkar_Abuzaid_Bailis_Zaharia_2018}.  Another issue that leads to the removal of cached data is the updation of raw files. Alpine tried to reduce reparsing for fresh data and updates happening on dataset by building indexes incrementally and refining subsets to accommodate updates \cite{Anagnostou_Olma_Ailamaki_2017}.  ReCache proposes to cache processed JSON or CSV raw data, which improves overall query processing time by 19-75\% \cite{Azim_Karpathiotakis_Ailamaki_2017}. However, the main memory caching techniques suffer from occasional reparsing issues and high main memory utilization for large datasets.

    S. Kim et al. have proposed to remove costly steps like sorting and redistribution while streamlining the conversion process to improve data loading in array-based DBMS SciDB \cite{Kim_Lee_Kim_Moon_2018}. However, best bulk loading techniques like COPY require 9x more resources to load the entire dataset into a database compared to raw storage \cite{Patel_Bhise_2019}. 
It has been observed that most application workloads frequently access very small part of the large dataset \cite{Stoica_Radu_Justin_J_Levandoski_2013}, \cite{Borovica_Gajic_Appuswamy_Ailamaki_2016}, \cite{Jain_Padiya_Bhise_2017}. A. Jain et al. have observed that 63\% of query workload can be answered using only 8\% of data \cite{Jain_Padiya_Bhise_2017}. Moreover, the smaller partition size reduced query execution time by 83\%. Loading frequently required data can reduce the overall workload execution time. However, some queries might require access to unloaded data. To solve this issue, hybrid systems have been developed consisting of in-situ engine and DBMS. The in-situ engine part helps in executing queries on unloaded raw data saving data loading time while the DBMS can reduce QET time for queries accessing frequently used data \cite{Patel_Yadav_Bhise_2022,Zhao_Cheng_Rusu_2015b}.

Invisible loading proposed loading processed data generated as a side effect of executing queries on raw datasets into column store DBMS to eliminate reparsing when main memory is full \cite{Abouzied_Abadi_Silberschatz_2013}. SCANRAW is one of the first hybrid systems which proposed monitoring CPU and IO to utilize their idle time to speculatively load additional data into DBMS in parallel to querying raw data using an in-situ engine \cite{Cheng_Rusu_2015}. The invisible loading and SCANRAW may process and load the entire dataset if workload queries access database attributes even once. Therefore, researchers developed cost-aware techniques which calculate the cost of accessing raw data files and DBMS to load only frequently accessed partitions of the dataset \cite{Patel_Yadav_Bhise_2022,Zhao_Cheng_Rusu_2015b}. The Query Complexity Aware (QCA) technique reduced the amount of loaded data by loading attributes required by complex queries while keeping attributes accessed by zero join simple queries in raw format \cite{Patel_Bhise_2022}. EEEQP \cite{kalgi_acids} optimizes hash joins through workload prediction, improving PCR (49\%), QET (8\%), and I/O efficiency (46\%) in edge-based smart city systems.

The techniques discussed earlier try to reduce data loading time. Partitioning datasets into smaller workload aware partitions reduces query execution time \cite{Padiya_Bhise_2017,Tang_Byna_Dong_Koziol_2020}. However, the increasing use of the cloud shifted the focus of researchers on resources used by each query to optimize the hardware resource utilization to reduce costs \cite{Pimpley_Li_Srivastava_Rohra_Zhu_Srinivasan_Jindal_Patel_Qiao_Sen_2021,Pimpley_Li_Sen_Srinivasan_Jindal_2022,Raza_Chrysogelos_Anadiotis_Ailamaki_2020}. QROP(Query and Resource Optimization) proposed considering the resource required by each query during query planning to reduce costs \cite{Viswanathan_Jindal_Karanasos_2018}. QROP ideology has been implemented using PCC (Performance Characteristic Curve) built using past resource utilization and query execution time data of repeating query \cite{Pimpley_Li_Sen_Srinivasan_Jindal_2022}. PCC helps allocate optimal resources to each query individually to reduce costs by trading off query response time. An elastic resource management technique has been proposed for an HTAP (Hybrid Transactional Analytic Processing) system, which trades-off OLTP throughput to reduce OLAP query time by providing more resources to OLAP queries \cite{Raza_Chrysogelos_Anadiotis_Ailamaki_2020}.

	The techniques and systems with features like optimizing resources utilized by data loading and query execution tasks, task scheduling, and query-specific resource allocation based on the real-time availability of hardware resources are rare. For example, Proteus \cite{Karpathiotakis_Alagiannis_Ailamaki_2016} is able to execute queries on heterogeneous data. It does not consider real-time utilization of resources nor allocates query-specific resources like PCC\cite{Pimpley_Li_Sen_Srinivasan_Jindal_2022}. Therefore, the following section discusses the proposed Resource Availability \& Workload aware Hybrid Framework (RAW-HF) to address those issues.	

\section{RAW-HF: Resource Availability \& Workload aware Hybrid Framework}


Many widely used Database Management Systems (DBMSs) lack real-time resource monitoring, automatic task scheduling, and query-specific resource allocation features. To address these limitations, this section introduces the Resource Availability \& Workload-aware Hybrid Framework (RAW-HF). RAW-HF aims to optimize resource utilization through workload-aware partitioning and efficient utilization of existing resources. Existing research suggests that hybrid systems, like in-situ engines with DBMSs, can achieve this objective \cite{Abouzied_Abadi_Silberschatz_2013,Cheng_Rusu_2015}. The in-situ engines with main memory indexing can query raw data immediately, reducing data to query time \cite{Alagiannis_Borovica_Branco_Idreos_Ailamaki_2012,Olma_Karpathiotakis_Alagiannis_Athanassoulis_Ailamaki_2020}. Additionally, loading of data into DBMS can help reduce query-to-result (QET) time for future queries \cite{Abouzied_Abadi_Silberschatz_2013,Cheng_Rusu_2015}. Therefore, the framework should be able to store and query raw data effectively, reducing data to first query time \& query execution time for hybrid systems.  

The existing cost-based dataset partitioning, task scheduling, and query-specific resource allocation techniques for hybrid systems require significant time to collect required data, train ML models, and automate different tasks \cite{Jindal_Interlandi_2021,Pimpley_Li_Sen_Srinivasan_Jindal_2022}. Therefore, this section discusses the proposed RAW-HF framework integrated with lightweight partitioning, task scheduling, and query-specific resource allocation algorithms. The following subsections discuss the architecture and modules of RAW-HF.

\subsection{RAW-HF Architecture} \label{RAW_HF_Architecture}

The RAW-HF framework comprises four modules, each dedicated to specific tasks. These modules include data loading, task scheduling, resource monitoring \& analysis, resource optimization by ensuring that only required data gets processed, and maximizing utilization of available resources to improve query performance.

\begin{itemize}

\item \textbf{Raw Data Query Processing (RQP) module:} 
This module processes application workloads automatically for hybrid systems. 
Such a combination of an in-situ engine and DBMS is chosen to build this hybrid system where any query can access data stored in DBMS and the raw data to answer a query when needed.


\item \textbf{Resource Monitoring (RM) module:} Monitoring resources used by each workload task is crucial.  This module interacts with external resource monitoring tools like \textit{top, htop} \cite{top, htop}, and \textit{iotop} \cite{iotop} to gather real-time hardware utilization data. 


\item \textbf{Optimizing Required Resources (ORR) module:} ORR optimizes resource utilization by processing only the necessary data. It combines features from two partial loading techniques: Query Complexity Aware (QCA) \cite{Patel_Bhise_2022}, and Workload \& Storage Aware Cost-Based (WSAC) \cite{Patel_Yadav_Bhise_2022} partitioning techniques. This approach reduces algorithm execution time compared to cost-based methods \cite{Zhao_Cheng_Rusu_2015b}. ORR aims to load only the essential attributes to minimize data loading time (DLT) and improve query execution time.


\item \textbf{Maximizing Utilization of Available Resources (MUAR) module:}  MUAR focuses on maximizing utilization of CPU and RAM resources during workload execution. It allocates query-specific work memory to manage resources effectively \cite{MUAR}. MUAR considers the real-time availability of resources to utilize resources efficiently and reduces total workload execution time by allocating additional \textit{work\_memory} compared to default configurations of DBMS. 

\end{itemize}

The interconnections between these modules and their interactions with external tools are illustrated in Figure \ref{Fig_2_RAW_HF_Architecture}. The RAW Data Query Processing module communicates with external DBMS and in-situ engines using a standard SQL-based Application Programming Interface (API). The Resource Monitoring module interacts with external resource monitoring tools to gather real-time hardware utilization information. This modular architecture allows RAW-HF to optimize resource utilization, reduce query execution times, and efficiently manage existing hardware resources for faster workload execution and cost savings. The next sections will provide more detailed insights into each module and their functionalities.


 \subsection{RAW-HF Modules}
\begin{figure}[htbp]
\centerline{\includegraphics[width=0.9\textwidth]{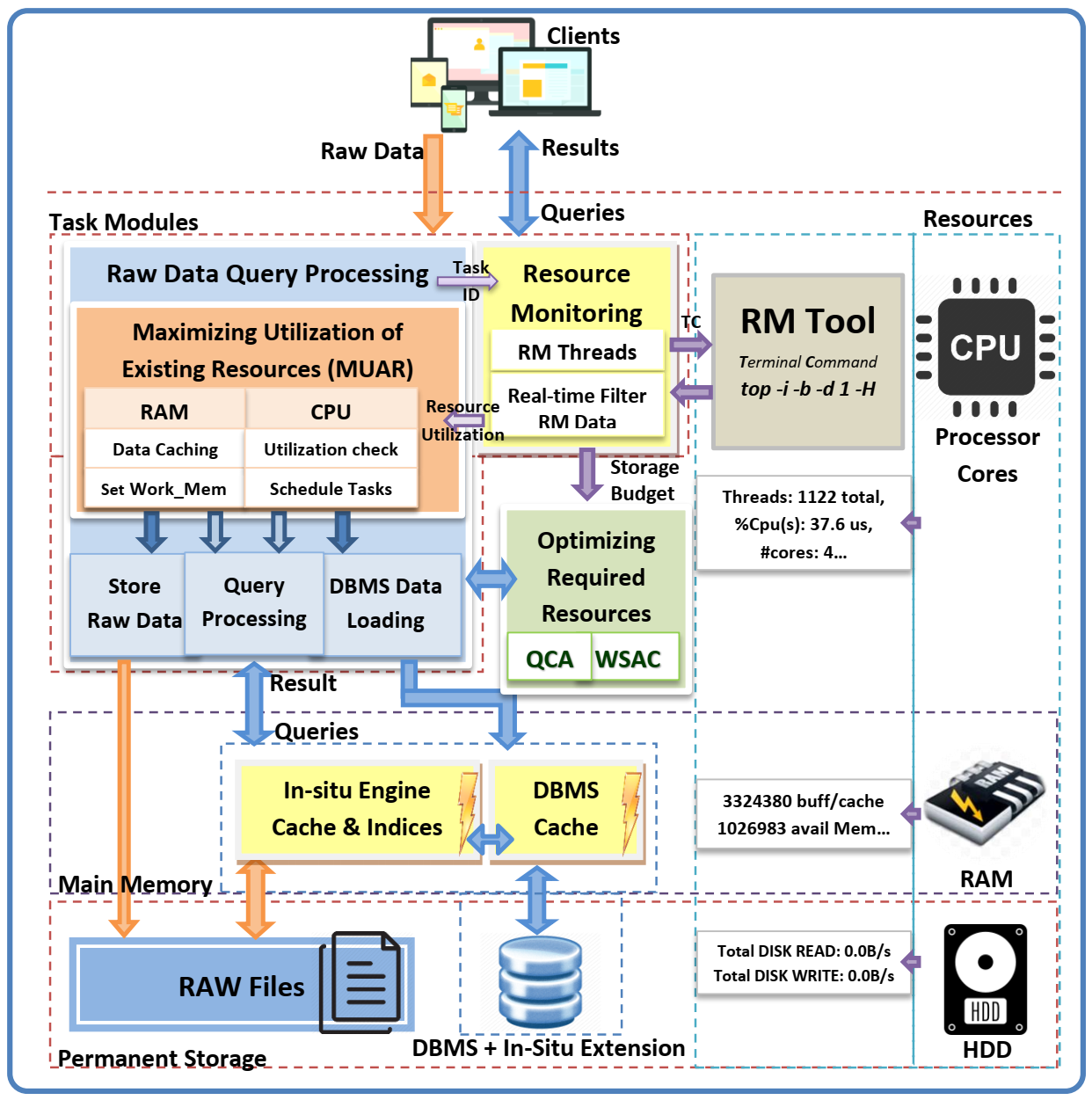}}
\caption{RAW-HF Architecture}
\label{Fig_2_RAW_HF_Architecture}
\end{figure}

 This section discusses the workings of each RAW-HF module and its algorithms in detail.
Figure \ref{Fig_2_RAW_HF_Architecture} illustrates the interconnections between all the RAW-HF modules, external tools, and how they interact with actual hardware resources.

 \subsubsection{Raw Data Query Processing (RQP) module: }
 The raw data query processing module of RAW-HF manages all the primary data and query processing tasks like storing data in CSV format, loading data in DBMS, and executing queries on raw or loaded data.  The framework must use the most effective ways of storing and loading datasets for hybrid systems. The invisible and speculative loading techniques propose to load data incrementally \cite{Abouzied_Abadi_Silberschatz_2013,Cheng_Rusu_2015}. However, loading vertical partitions incrementally takes more time and resources due to dataset partitioning, data loading, and creating new or updating the existing data table steps. 
 Basic experiments and research have determined that the fastest method is COPY command for disk-based databases, which loads data from CSV (comma-separated value) file \cite{Dziedzic_Karpathiotakis_Alagiannis_Appuswamy_Ailamaki_2017,Patel_Bhise_2023a}. Experiment results also concluded that parallel loading could not improve DLT time for disk-based permanent storage devices. 
 While storing data in main memory using RAM Files System (RAMFS) with parallel loading can reduce DLT by 20-30\%. However, COPY \& COPY with RAMFS takes 6x or more time than storing data in raw format on disk \cite{Patel_Bhise_2023a}. Therefore, the RQP module performs the below tasks on data received from different sources to reduce time and resource utilization.
 \begin{itemize}
     \item \textbf{Store Raw Data:} This sub-module stores received data in raw data files (CSV) to reduce initial data storage time. These raw data files must be linked to the in-situ engine to execute queries directly. 
 It creates the raw file partitions based on the ORR module input to RQP, to reduce partition creation and re-partitioning costs. 

  \item \textbf{Query Processing:} This sub-module needs to execute initial queries on raw data files, reducing data to first query time. 
 It also utilizes the DBMS loaded data to answer queries faster. This means it supports multi-format query processing. For example, if any query requires some data from raw format and some from DBMS, then this framework can handle such cases as well without requiring all data in a single format. This helps when workload-aware partitions are created based on ORR input, but some ad-hoc queries require execution. 
 
 \item \textbf{DBMS Data Loading:} 
RQP manages raw file partitioning, table creation, and data loading steps based on ORR module input.
 \end{itemize}

 \subsubsection{Resource Monitoring (RM) module: }
 Researchers have observed that knowing the resources used by a query beforehand can help in choosing the best query plans to optimize resource utilization and costs \cite{Li_Wang_Wang_Sun_Peng_2022,Pimpley_Li_Sen_Srinivasan_Jindal_2022,Zafeiropoulos_Fotopoulou_Filinis_Papavassiliou_2022}. The RM module takes care of monitoring real-time utilization of hardware resources. It associates task ID to resources utilized by each task and filters unwanted data for faster offline or online analysis. RM tries to impose the minimal overhead of monitoring and filtering query-specific resource utilization data. 
The two sub-modules of RM module and their tasks have been discussed below.

 \begin{itemize}

     \item	\textbf{RM Threads:} Most DBMSs do not natively support resource monitoring. Therefore, task of this sub-module is to interact with external resource monitoring tools. It initiates resource monitoring tasks and receives the overall \& per-process resource utilization data. This module creates a resource monitoring thread and executes terminal commands on multiple external tools like \textit{top} and \textit{iotop}, as one tool may not provide all necessary data. 
     It also filters out the required data from hundreds of systems, DBMS, in-situ, and other processes running simultaneously. 
     The modified version of earlier proposed RM module has been used for RAW-HF, which can identify resource utilization data of each individual workload task from \textit{top \& iotop} tools. Most cloud service providers also do not provide resource utilization data for each query separately. This data allows MUAR module of RAW-HF to allocate accurate resources to repeating queries and efficiently utilize available resources. 

     \item 	\textbf{Real-Time Filter:} 
     This sub-module filters out additional unwanted data to reduce the size of output files for offline analysis. Basic experiments have shown that a few GBs of data get generated by monitoring resources with 0.1 to 1sec frequency within couple of hours. 
     Filtered data is stored in CSV file format for faster storage, imposing minimal overhead. The real-time availability of the overall system resources is also filtered and stored in shared variables. These data are used by MUAR in real time for task scheduling and memory allocation decisions. The real-time availability of CPU and RAM resources is not stored in CSV files every second to reduce IO.
    
 \end{itemize}

\subsubsection{Optimizing Required Resources (ORR) module: }
This module of RAW-HF tries to optimize resource utilization by processing only required data. A lightweight query complexity, workload, storage cost, read cost, and write cost-aware partial loading technique has been developed to achieve this goal. 
ORR combines features of two partial loading techniques: 1) Query complexity aware (QCA) \cite{Patel_Bhise_2022}, and 2) Workload and Storage aware Cost-based Technique (WSAC) \cite{Patel_Yadav_Bhise_2022}.
 
\begin{figure}[htbp]
\centerline{\includegraphics[width=\textwidth]{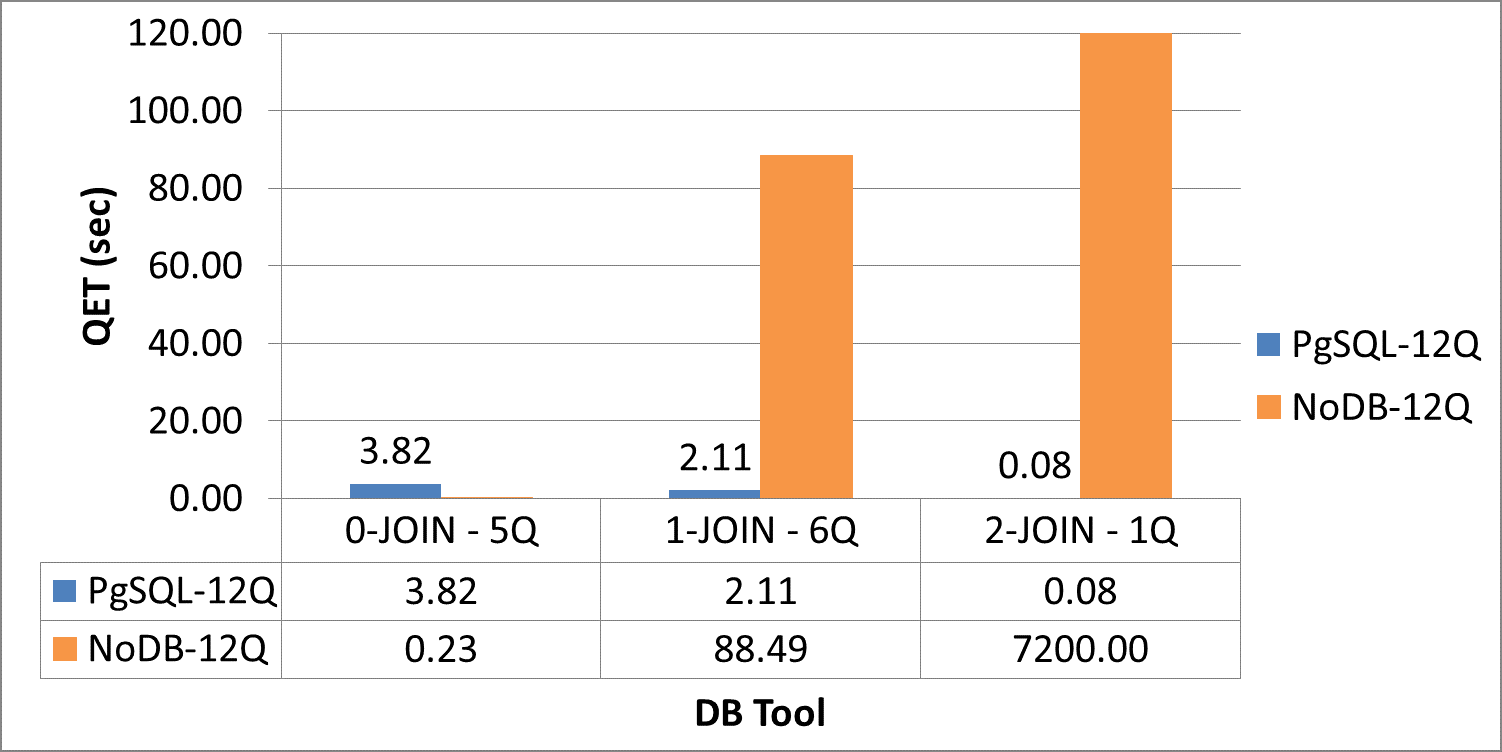}}
\caption{SDSS: Query Classification based on Join Count}
\label{Fig_32_SDSS_Query_Classification_based_on_Join_Count}
\end{figure}

The proposed workload and query complexity aware algorithms of ORR use lightweight query identification and partitioning steps of QCA\cite{Patel_Bhise_2022} and WSAC\cite{Patel_Yadav_Bhise_2022} to reduce algorithm execution time (AET) compared to cost-based workload aware partial loading technique (WA) \cite{Zhao_Cheng_Rusu_2015b, Patel_Yadav_Bhise_2022}. The idea behind the QCA technique is to partition the dataset and distribute the workload in such a way that queries performing faster on a given tool can be allocated to that tool \cite{Patel_Bhise_2023a}. QCA algorithm had identified the type of queries best suited for a given tool using initial RQP results \cite{Patel_Bhise_2022} as shown in Figure \ref{Fig_32_SDSS_Query_Classification_based_on_Join_Count}. The analysis has shown that zero-join queries perform faster in raw engines than in traditional DBMS. While queries having multiple joins are slow in raw engines. Therefore, the technique classified the query workload into two query types. The first type is for simple queries (SQ), which contain zero join. The second query type included the remaining one or more join queries. This second category of queries is called complex queries (CQ) in this paper.  

ORR proposes to utilize the above-mentioned observation to reduce the amount of data being loaded into DBMS. Similar to QCA, ORR also eliminated DLT for attributes used by SQ by loading only CQ attributes. ORR algorithm can be divided into three parts, 1) Query Complexity Identification QCI, 2) Grouping of Attributes based on query classification GRA, \& 3) Queries Coverage (QC). 

\begin{algorithm}[htbp]
\caption{ORR Module: Optimization of Required Resources}
 \texttt{\textbf{Data}:{ \textit{w\_l}} = Workload List;
 \textit{QT}  = Query Types Dictionary;		  
 \textit{q\_l} = Query List;
 \textit{B} = Storage budget B in MB; 
 \textit{que\_d} = Dictionary of Queries;
 \textit{s\_d}  = Schema Dictionary;
 \textit{QT\_P} = Query Type Partitions; 
 \textit{QT\_P’}= Final Query Type Partitions for given budget B;
 \textit{q\_l}  = Query List;
 \textit{PCQ}  = Partially Covered Queries List;				  	  
 \textit{ca\_l} = List of covered Attributes; 
  \textit{rqa\_l} = List of Remaining Query Attributes;
 \textit{cq\_l} = List of covered Queries;
\\ \textbf{Result}: SQ-Raw, CQ-DB \& CAP Partitions;
\\ \rule{\textwidth}{0.9pt}
\\ \textcolor{white}{{\#}}\qquad \textcolor{M_green}{\qquad	 \# Query Complexity Identification}
\\ 1. \textbf{def} \textbf{\textit{QCI}}(\textit{w\_l}, \textit{que\_d}, \textit{s\_d}):
\\ 2. \qquad	 \textbf{For} each task T in \textit{w\_l} do 
\\ 3. \qquad\qquad			\textbf{If} T.Statement has multiple tables
\\ 4. \qquad\qquad\qquad			\textit{	QT[T.Q\_ID]} = 1
\\ 5. \qquad\qquad			\textbf{Else}
\\ 6. \qquad\qquad\qquad			\textit{	QT[T.Q\_ID]} = 0
\\ 7. \qquad	 \textbf{End}
\\ 8. \qquad	 Get \textit{QT\_P0, QT\_P1 }= \textbf{\textit{GRA}}(\textit{que\_d}, \textit{QT})
\\ 9.	\textbf{Return} final partitions \textit{QT\_P’}; \textcolor{M_green}{\qquad	 \#Return all QT partitions}
\\ \rule{\textwidth}{0.9pt} 
\\ \textcolor{white}{{\#}}\qquad \textcolor{M_green}{\qquad					\#Grouping of Attributes}
\\ 10.	\textbf{def GRA}(\textit{que\_d}, QT)
\\ 11.\qquad	 \textbf{For} each query i in \textit{que\_d}:
\\ 12.\qquad\qquad		\textbf{For} each attribute j in \textit{que\_d}[i]:
\\ 13.\qquad\qquad\qquad \textbf{If} \textit{QT[i]} == 0
\\ 14.\qquad\qquad\qquad\qquad Add \textit{j} in\textit{ QT\_P0}
\\ 15.\qquad\qquad\qquad\qquad\qquad \textbf{If} budget \textit{B} is limited
\\ 16.\qquad\qquad\qquad\qquad\qquad\qquad	 					\textit{QT\_P’} = \textbf{QC}(\textit{i, ca\_l, cq\_l, B})
\\ 17.\textcolor{M_green}{Repeat above step until PCQ is empty for different B}
\\ 18.\qquad\qquad\qquad \textbf{Else}
\\ 19.\qquad\qquad\qquad\qquad Add \textit{j} in \textit{QT\_P1}
\\ 20.\qquad\qquad	  	\textbf{End}
\\ 21.\qquad	 \textbf{End} 
\\ 22.	\textbf{Return} \textit{QT\_P0, QT\_P1,QT\_P’};
\\ \rule{\textwidth}{0.9pt}
\\ \textcolor{white}{{\#}} \textcolor{M_green}{\qquad	  \#Grouping of Attributes based on Budget B}
\\ 23. \textbf{def QC}(\textit{i, ca\_l, cq\_l, B})
\\ 24.\qquad	 	\textbf{If} (SUM(size of all attributes of query que\_d[i]))< B
\\ 25.\qquad\qquad	 		\textbf{For} each attribute A in que\_d[i]
\\ 26.\qquad\qquad\qquad \textbf{If} A is not in \textit{ca\_l} \& size of A<remaining budget \textit{B}
\\ 27.\qquad\qquad\qquad\qquad	 				Add A in \textit{rqa\_l} list
\\ 28.\qquad\qquad	 		\textbf{If} size of \textit{rqa\_l} < remaining budget \textit{B}
\\ 29. \qquad\qquad\qquad Move all attributes of \textit{rqa\_l} in \textit{ca\_l} \& update \textit{B}
\\ 30. \qquad\qquad\qquad Add query q in \textit{cq\_l}
\\ 31.\qquad\qquad	 		\textbf{Else}
\\ 32.\qquad\qquad\qquad	 			Add query q in PCQ 
\\ 33.	\textbf{Return} \textit{ca\_l, cq\_l, rqa\_l};}
\end{algorithm}


The Query Complexity Identification (QCI) function of QCA\cite{Patel_Bhise_2022} technique first identifies the Simple Query (SQ) and complex query (CQ) type queries stored in the workload list. The algorithm uses the simple logic of counting no. of tables present in the query statement. The query is classified as a complex query; if two or more table instances are found in the query statement.
It also identifies the hot data using workload information and tries to cover most queries for the given storage budget. The cost function finds the size of each workload attribute to decide if that attribute should be loaded into the database or not for the given storage budget B. Contrary to the HTAP systems, ORR tries to achieve faster query execution times with minimal replication and loaded data partitions.

\begin{table}[htb]
\caption{Query Type Dictionary (QT)}
\label{Table_8_Query_Type_Dictionary}
\begin{tabular}{|P{0.4\linewidth }|P{0.04\linewidth }|P{0.04\linewidth }|P{0.03\linewidth }|P{0.03\linewidth }|P{0.04\linewidth }|P{0.04\linewidth }|P{0.04\linewidth }|P{0.04\linewidth }|P{0.04\linewidth }|P{0.04\linewidth }|P{0.04\linewidth }|}
\hline
\textbf{Key (Q\_ID)}       & 1 & 2 & 3 & 4 & 5 & 6 & 7 & 9 & 10 & 11 & 12 \\ \hline
\textbf{Value(Query Type)} & 1 & 0 & 1 & 0 & 1 & 0 & 0 & 1 & 0  & 1  & 1  \\ \hline
\end{tabular}
\end{table}

Table \ref{Table_8_Query_Type_Dictionary} shows the query ID and query complexity type updated in QT as key-value pair. 
The single table instance queries are classified as simple queries SQ. The GRA function groups the SQ and CQ attributes in two different lists \textit{QT\_P0} and \textit{QT\_P1}. 
The intersection of these two lists provides the list of common attributes partition CAP. 
SQ partition can be stored in raw format, while the CQ partition needs to be loaded in DBMS similar to QCA \cite{Patel_Bhise_2022}. 
However, ORR further refines the partitions based on storage budget B in QC function. 
These steps reduce partition size and find new group of queries covered by smaller partitions when storage budget B is limited and smaller than initial QT(P0 or P1) partition size. For most cases, the first round of partitioning might be enough. Otherwise, QC function needs to be called until all attributes are covered for different storage budgets for 
distributed systems like WSAC \cite{Patel_Yadav_Bhise_2022} to obtain multiple smaller partitions for each system. Most frequent queries and storage budget list can be sorted in descending order to reduce iterations \& cover frequent queries first. 

\begin{algorithm}[htbp]
\caption{MUAR Module: Maximizing Utilization of Available Resources}
 \texttt{\textbf{Data}: \textit{w\_l} = Workload List; 
\\\textit{RM\_AR} = Real-Time availability of CPU, RAM, \& IO resources in \%; 
\\\textit{TR} = Total RAM;
\\ \textit{Min\_RR} = Minimum resources  required to schedule a task;
\\ \textit{J\_C} = Join Count of a query q; P\_C = Process count that free CPU can handle;\\ \textit{WM\_value} = Work memory value; CPU\_C = CPU cores count of experiment machine;
\\\rule{\textwidth}{0.9pt}
\\ 1. \textbf{def} \textbf{\textit{MUAR}}(\textit{w\_l, RM\_AR, Min\_AR}):
\\ 2.\qquad		\textbf{For} each query \textit{q} in \textit{w\_l}:
\\ 3.\qquad\qquad			\textbf{while} resources are not available-\textit{RM\_AR$<$Min\_AR}:
\\ 4.\qquad\qquad\qquad					sleep 0.1sec
\\ 5.\qquad\qquad			\textbf{if} minimum resources are available-\textit{RM\_AR$>$Min\_AR}:
\\ 6.\qquad\qquad\qquad					Set work\_memory = \textit{WM\_Query} (\textit{RM\_AR, q})
\\ 7.\qquad\qquad\qquad					Add a new query thread in parallel.
\\ 8.\qquad		\textbf{End} \textbf{for} 
\\ 9. \textbf{Exit};
\\\rule{\textwidth}{0.9pt}
\\ 10. \textbf{def} \textbf{\textit{WM\_Query}} (\textit{RM\_AR, q})
\\ 11.\qquad		\textit{ J\_C }= Count Joins in a query q
\\ 12.\qquad	 	\textit{ P\_C} = \textit{RM\_AR.CPU/(100/CPU\_C)}
\\ 13.\qquad			\textit{ WM\_value =(((RM\_AR.RAM/P\_C)*(TR/CPU\_C))*(J\_C/4.0))};
\\ 14. \textbf{Return} \textit{WM\_value};}
\end{algorithm}

 \subsubsection{Maximizing Utilization of Available Resources (MUAR) module: }
 This module implements the Maximizing Utilization of Available Resources (MUAR) algorithm to improve the utilization of existing resources. It tries to maximize CPU and RAM resource utilization automatically during workload execution. MUAR considers real-time resource monitoring values of CPU, RAM, and IO resource utilization stored in the global structure \textit{RM\_AR}.
 \textit{RM\_AR} is continuously updated by the resource monitoring module.
  MUAR adds a new task for processing if all three values of \textit{RM\_AR} are greater than the minimum required CPU, RAM \& IO resources stored in \textit{Min\_RR}. Before executing the query in a new thread, the \textit{WM\_Query} function sets the work memory for each complex query to increase RAM utilization. The \textit{WM\_Query} function first counts the number of joins used in the given query and stores the count in \textit{J\_C}.

  The \textit{WM\_Value} in line 18 calculates the work memory value for new queries based on available memory \textit{RM\_AR.RAM}, process count (\textit{P\_C}), total RAM (\textit{TR}), and join count \textit{J\_C}. The first part of the equation divides the available RAM between the maximum processes that the available CPU cores can handle. The second part defines the maximum RAM that can be assigned to a thread, while the third part helps in allocating more RAM to complex queries considering join count \textit{J\_C}. MUAR also tries to estimate required work memory considering previous \textit{work\_mem}, disk writes, current \& past record count ratio for frequent queries to achieve the best QET time. 
 Whenever required work memory exceeds the available memory, 90\% of available memory is allocated to achieve the near best QET.
\section{Experimental Setup}

Experimental setup details like hardware-software setup, dataset, query set, \& exp. flow are discussed in this section.   

\subsection{Hardware \& Software Setup}
The experimental machine uses a quad-core Intel i5-6500 CPU clocked at 3.20GHz. It has 16GB of RAM. The operating system of the machine is running a 64-bit Ubuntu 18.04 LTS. The machine has a 500GB SATA hard disk drive to store raw datasets and DBMS databases. The disk rotation speed is 7200RPM. The robust RAW-HF framework has been developed by modifying \& integrating raw data query processing \cite{Patel_Bhise_2019}, resource monitoring\cite{Patel_Bhise_2023a}, and MUAR\cite{MUAR} frameworks developed earlier as modules with ORR proposed in this paper. The framework uses Eclipse to run Java code. It uses state-of-the-art open-source DBMS PostgreSQL as \textit{work\_mem} can be configured at runtime and NoDB in-situ engine with processed data caching capability. The source code of NoDB (PostgresRAw) is available on Github \cite{GitHub_HBPMedical_PostgresRAW}. Linux command line tools \textit{top} \cite{top} and \textit{iotop} \cite{iotop} provide real-time resource utilization data of CPU, RAM \& IO resources to the RM module. RAW-HF source code is also published on Github \cite{raw_hf_git}. 

\subsection{Dataset \& Query Set}
RAW-HF performance has been tested using two real-world datasets known as  Sloan Digital Sky Survey (SDSS) \cite{DR16_Ahumada2020} and Linked Observation Data (LOD)  \cite{knoesis_2010,Padiya_Bhise_2017} \cite{DR16_Ahumada2020}. 8GB partition with 35M records  containing descriptions of blizzard and hurricane observations has been extracted from LOD dataset. It is a benchmark RDF dataset used to investigate the performance of the MUAR module of RAW-HF. Data release 16 of SDSS has been used to check the ORR and MUAR phases of RAW-HF. The LOD dataset has a single narrow table with only three columns subject, object, and predicate. Therefore, vertical partitioning techniques used in ORR phase cannot be used for LOD dataset to improve WET. 16 standard RDF queries having different numbers of joins are used for experiments. Nine queries out of 16 queries of LOD workload have 5 joins. Queries with multiple self-joins have been considered because these queries need to process complex join operations, which require significant resources. On the other hand, SDSS query workload has only one query with two joins, while the remaining 11 queries have zero or one join. 19GB partition having 4M records of \textit{PhotoPrimary} view has been used to represent SDSS dataset, because 55\% of SDSS Dr-16 query workload used the \textit{PhotoPrimary} view. The extracted 12 query workload represents 51\% of the entire workload.

\subsection{Implementation Block Diagram} 

\begin{figure}[htbp]
    \centering
    \includegraphics[width=\textwidth]{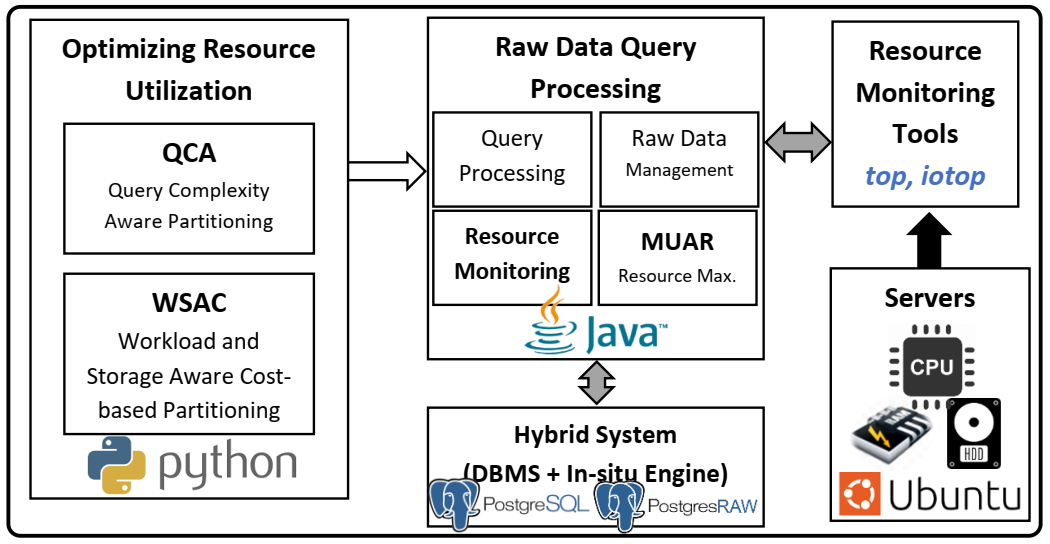}
    \caption{Implementation Setup Diagram}
    \label{Fig_13_Implementation_Setup_Diagram}
\end{figure}

This section provides implementation setup detail of each RAW-HF component. Figure \ref{Fig_13_Implementation_Setup_Diagram} shows the external tools and language names used to implement the entire framework. It can be seen that QCA and WSAC are implemented using Python. Python is easy to code and provides a rich library of functions that reduces lines of code to develop complex algorithms. The framework is implemented using Java code because Java is generally faster and more efficient than Python. Java is a compiled language. It reduces algorithm execution time (AET) for real-time algorithms compared to Python. Therefore, time bound real-time algorithms like resource monitoring and MUAR have been implemented using Java \cite{PATEL2024100643}.

The RAW-HF uses state-of-the-art DBMS PgSQL to handle database tasks. PgSQL is also open source. The in-situ engine used here is the NoDB \cite{Alagiannis_Borovica_Branco_Idreos_Ailamaki_2012}. NoDB is one of the first few in-situ engines which allows querying raw data using SQL and indexes processed data in main memory. NoDB was developed for Human Brain Project and is available on GitHub \cite{GitHub_HBPMedical_PostgresRAW}. The \textit{top} and \textit{iotop} resource monitoring tools have been used to monitor the resources utilized by Ubuntu system and other processes.

\section{Evaluation Parameters} \label{s_8_3_evaluation_parameters}
This section discusses all the evaluation parameters used in Section \ref{s_9_results} to compare experimental results. The parameters have been grouped into two categories: 1) Input Parameters, and 2) Output Parameters.

\subsection{Input Parameters}
The following list of parameters has been provided as input to different algorithms of RAW-HF. 

\subsubsection{Workload Files} 
The workload parameters like schema files and list of workload queries help develop workload-aware techniques like oRR (QCA + WSAC) and MUAR of the proposed RAW-HF framework. 





\subsubsection{Resource Utilization Parameters} 
Resource utilization parameters represent the percentage of CPU, RAM, or IO (Disk read/writes in MB) resources utilized during query execution. Different algorithms of RAW-HF have used the historical or real-time values of these parameters.  




\subsubsection{Work\_mem (MB):} It is PostgreSQL configuration parameter. The size of RAM allocated to each workload query has been configured using this Work\_mem parameter in real-time to improve QET.




\subsection{Output Parameters}
The following list of output parameters helps us identify the reductions or improvements achieved by applying proposed techniques compared to existing tools \& techniques.

\subsubsection{Query Performance Parameters} 
These parameters display the time taken by in-situ engines or DBMS to complete given workload tasks. For example, QET (sec), DLT (sec).  

 

\subsubsection{Workload Performance Parameters:} Workload Execution Time (WET) is the total time system takes to execute a given workload. For single-thread execution, WET can be simply calculated by summing DLT and QET. However, in multi-thread execution of workload, the total time is calculated by subtracting the experiment end time from the start time. \begin{equation}
     \textbf{WET = DLT + QET}
\end{equation}.

\subsubsection{Fraction of Attributes Accessed (FAA)
 (\%):} This parameter shows the number of attributes accessed by the workload queries. 
\begin{equation}
     \textbf{FAA = No. of Accessed Attributes / Total Attributes}
\end{equation}.

\subsubsection{Fraction of Attributes Loaded (FAL)
 (\%):} This parameter shows the number of attributes loaded into DBMS by the technique. 
\begin{equation}
     \textbf{FAL = No. of Attributes Loaded / Total Attributes}
\end{equation}.
 
\subsection{Resource Utilization Parameters} 
These parameters represent the percentage of CPU, RAM, or IO resources utilized to execute a given workload. 






\section{RAW-HF Results}\label{s_9_results} 
This section presents the results of RAW-HF after combining all techniques proposed in ORR \& MUAR Phases. Sections \ref{5_1_WET} and \ref{5_2_RU} present results obtained using SDSS dataset. RAW-HF performance is also compared with state-of-the-art tools and techniques based on the WET and resource utilization parameters.

\subsection{Optimizing Required Resources (ORR)  }
\label{ORR_SDSS}

This section presents the ORR results applied to SDSS dataset. ORR phase tries to optimize resource utilization by processing only required data. SDSS is a broad table dataset, which means most of the table attributes may not get accessed by the workload queries. ORR identifies such attributes and saves DLT time by not processing such attributes. ORR phase cannot be applied to LOD dataset, because \textit{LODTriples} table had only three attributes. All of these attributes are accessed by most of the workload queries. Therefore, vertical partitioning used by ORR cannot be applied. Hence, the ORR phase can help reduce WET for broad table datasets only. In future, horizontal partitioning or hybrid techniques like DWAHP \cite{Padiya_Bhise_2017} can be incorporated to improve WET for narrow table datasets.

\begin{figure}[!htbp]
    \centering
    \includegraphics[width=\textwidth]{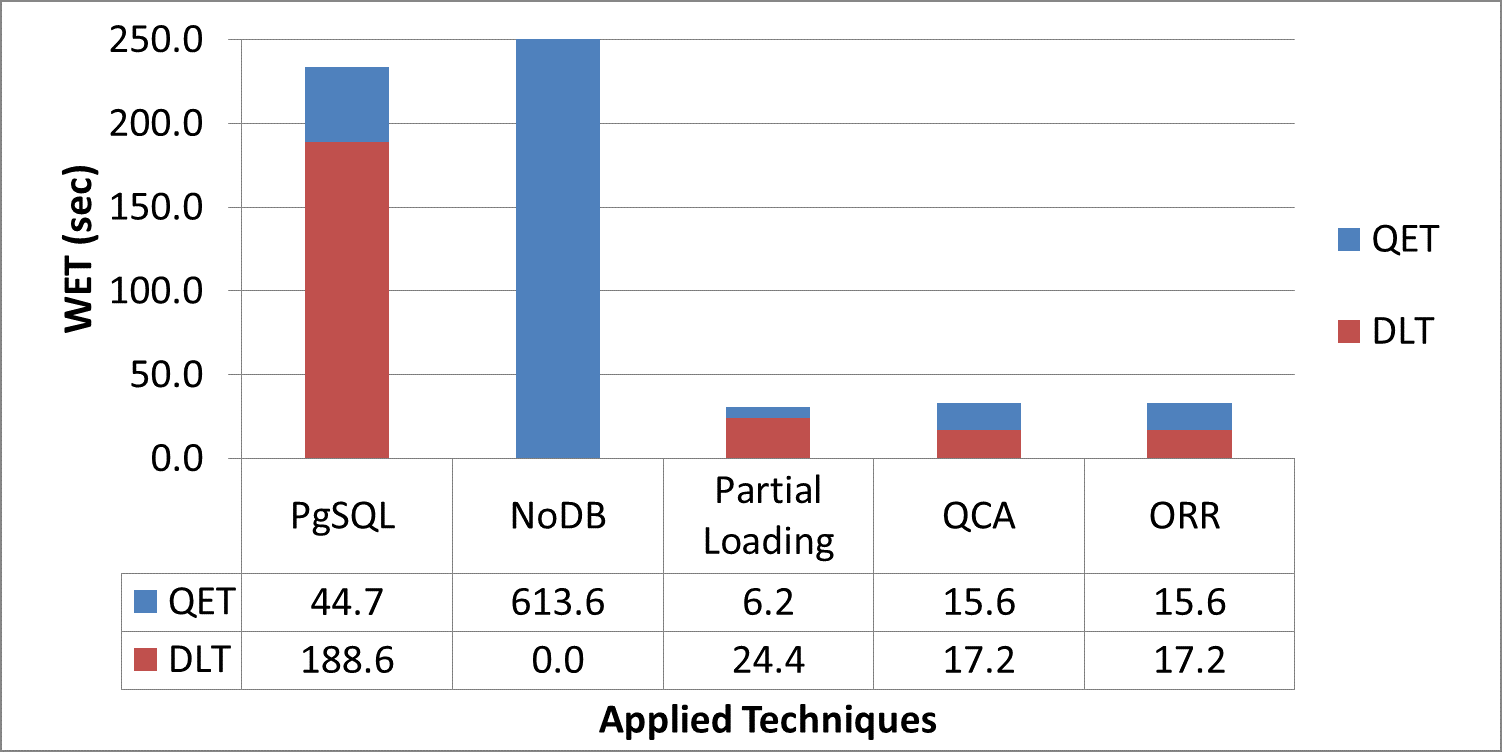}
    \caption{ORR: WET comparison for SDSS dataset}
    \label{Fig33_1_ORR_WET}
\end{figure}

Figure \ref{Fig33_1_ORR_WET} displays the WET time taken by in-situ engine NoDB(PostgresRAW), PostgreSQL DBMS, Workload Aware state-of-the-art partitioning technique(WA), QCA\_Case-V, and the ORR. ORR uses the case CASE-V result of QCA proposed cases because it incorporates replication of required attributes for multi-core execution. There was enough memory budget to store all the complex query partition attributes on a single machine. Therefore, both QCA and ORR phase output partitions and results are similar. However, ORR phase improved algorithm ensures the replication of required attributes and the creation of smaller partitions for a distributed environment when enough memory is not present on a single node. The results have been obtained by executing data loading and workload query tasks sequentially using a single CPU core only. 
It can be seen that WA, QCA, and ORR perform similarly on single CPU core execution. However, Figure \ref{Fig_42_WET_Comparison} shows that the ORR chosen partitions help surpass WA performance when utilizing all available CPU cores. Here, the ORR phase reduced the WET for the SDSS dataset by 94\% and 86\% compared to NoDB and PostgreSQL DBMS, respectively. 

\subsection{MUAR } \label{MUAR_for_Different_Datasets}
This section analyzes the experimental results after applying MUAR on datasets like SDSS and LOD. MUAR tunes parameters like  \textit{work\_mem} in real-time for PostgreSQL DBMS. 

\begin{figure}[htbp]
    \centering
    \includegraphics[width=\textwidth]{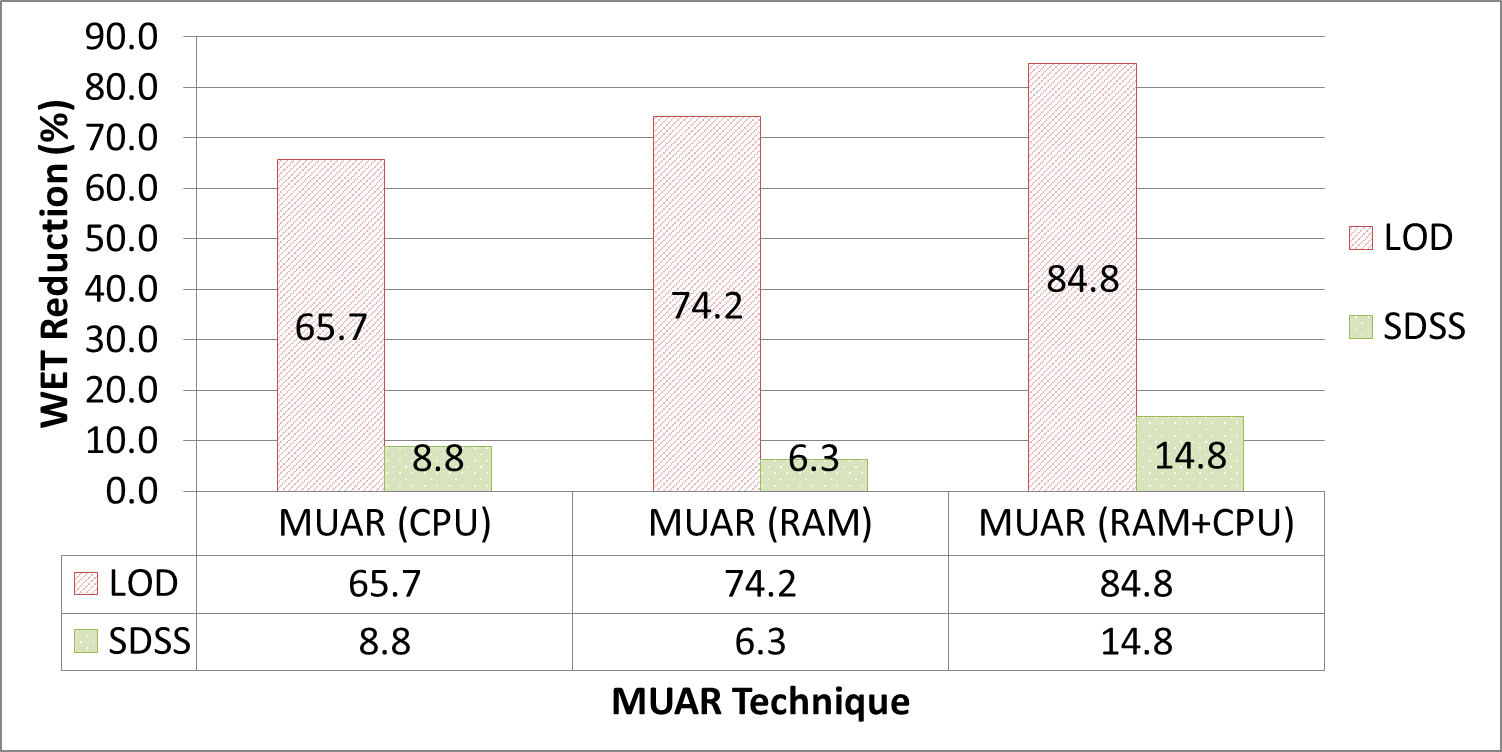}
    \caption{Impact of MUAR on WET } 
    \label{Fig44_MUAR_for_Different_Datasets}
\end{figure} 

Figure \ref{Fig44_MUAR_for_Different_Datasets} shows the impact of RAM and CPU maximization techniques used by MUAR on LOD and SDSS datasets. It can be seen that individual and combination of resource maximization techniques used by MUAR are more effective on the LOD dataset. It is visible that the WET reduced for LOD dataset is 5.7 times more compared to the SDSS dataset. The large difference in WET improvement is due to the different characteristics of both datasets. The \textit{PhotoPrimary} table of SDSS contains 509 attributes, while the LOD dataset is a narrow table dataset with only three attributes. For SDSS, loading all 509 attributes in the dataset required 188.63sec while QET time is only 44.7sec. For SDSS, 80.8\% of WET is spent loading data into DBMS due to the 4.7GB size per 1M records. While for LOD dataset only spent 0.4\% of the WET time loading data. MUAR is not using the IO maximization technique. Therefore, DLT time cannot be improved using CPU \& RAM maximization techniques. This means the effect of CPU and RAM maximization techniques on WET depends on QET time reduction only. For the LOD dataset, 99.6\% of the time is spent on query execution. In comparison, SDSS QET time is less than 20\% of WET. Therefore, executing queries in parallel helps reduce overall WET by only 8.8\% for SDSS. On the other hand, 99.6\% of the workload can be executed in parallel for the LOD dataset. Therefore, executing queries in parallel achieved 65.7\% reduction in WET for the LOD dataset workload.

The RAM maximization technique used by MUAR allocates more work memory to complex queries to reduce QET. The allocation of more work memory helps complex multi-join queries execute faster as disk access reduces significantly. For the LOD dataset, 56\% of the query workload had five join queries, while 87\% had two or more joins. On the other hand, query workload of the SDSS dataset had less than 1\% of queries with two joins. Therefore, MUAR RAM maximization techniques also achieved better results in reducing QET for the LOD dataset than SDSS. It can be seen in Figure \ref{Fig44_MUAR_for_Different_Datasets} that MUAR RAM maximization achieved 11 times more reduction in WET than the LOD dataset. For SDSS, allocating more RAM did not help reduce QET due to a simpler query workload with fewer joins, as disk writes were already less or non-existent. However, caching the entire dataset into main memory helped reduce overall WET by 6.3\% for the SDSS workload. Therefore, MUAR(CPU+RAM) is more effective for datasets having complex query workloads like LOD.

\subsection{RAW-HF }\label{5_1_WET} 
ORR partitions the dataset to optimize resource utilization. While, MUAR is task scheduling and resource allocation module. RAW-HF combines both by utilizing ORR partitions to reduce unnecessary processing of data during query execution, while RAW-HF uses MUAR module to execute those queries in parallel and allocate more resources to complex queries. Most existing systems either employ optimization or resource maximization techniques. RAW-HF employs both techniques, ensuring applicability to most real-world workload requirements. 

\subsubsection{Workload Execution Time } 

\begin{figure}[htbp]
    \centering
    \includegraphics[width=\textwidth]{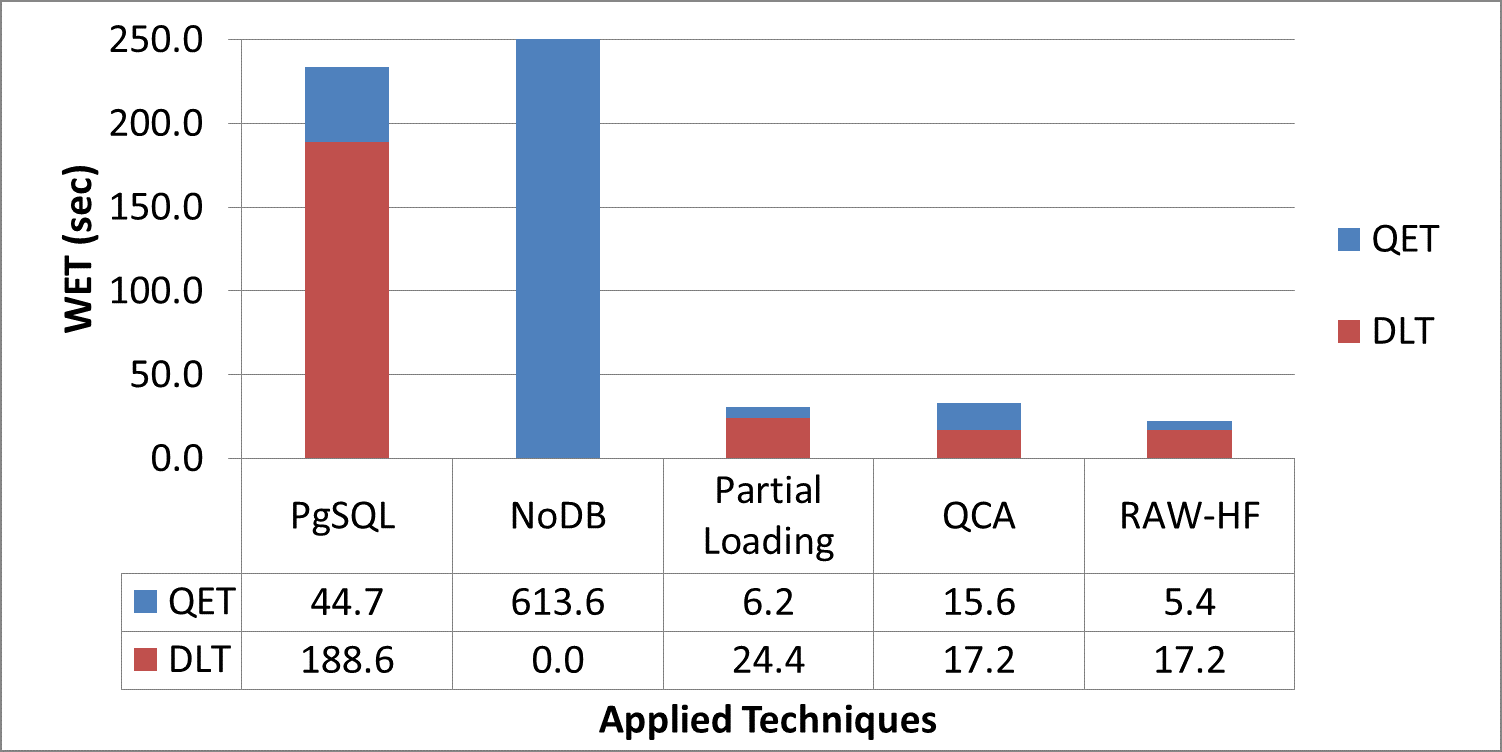}
    \caption{WET: Comparison}
    \label{Fig_42_WET_Comparison}
\end{figure} 

Figure \ref{Fig_42_WET_Comparison} shows the comparison of all the techniques with RAW-HF. The first column shows the WET time required by traditional DBMS PgSQL. 2nd column shows the raw data query processing time required by NoDB, which does not require loading any data into a DBMS-specific structure. 
The ORR combines the best features of QCA \&  WSAC technique to reach the workload execution time of 32.8 sec, which is still 2.2 sec more compared to workload aware Partial Loading technique \cite{Zhao_Cheng_Rusu_2015b}. However, 
RAW-HF completes the execution of the given workload within 22.6 sec compared to the 30.6 sec required by the Partial Loading technique \cite{Zhao_Cheng_Rusu_2015b}. It can be observed that the RAW-HF achieved a total reduction of 90.31\%, 96.32\%, and 26.14\% compared to PgSQL\cite{PostgreSQL}, NoDB\cite{Alagiannis_Borovica_Branco_Idreos_Ailamaki_2012}, and workload-aware partial loading technique\cite{Zhao_Cheng_Rusu_2015b} by combining techniques implemented in ORR \& MUAR modules. RAW-HF benefits from low DLT time achieved by only loading attributes used by complex queries in the ORR module for SDSS. 
Additionally, simple queries complete execution in parallel to data loading tasks utilizing available resources efficiently with the help of MUAR. 

\subsubsection{RAW-HF:	Resource Utilization} \label{5_2_RU}
\begin{figure}[htbp]
    \centering
    \includegraphics[width=\textwidth]{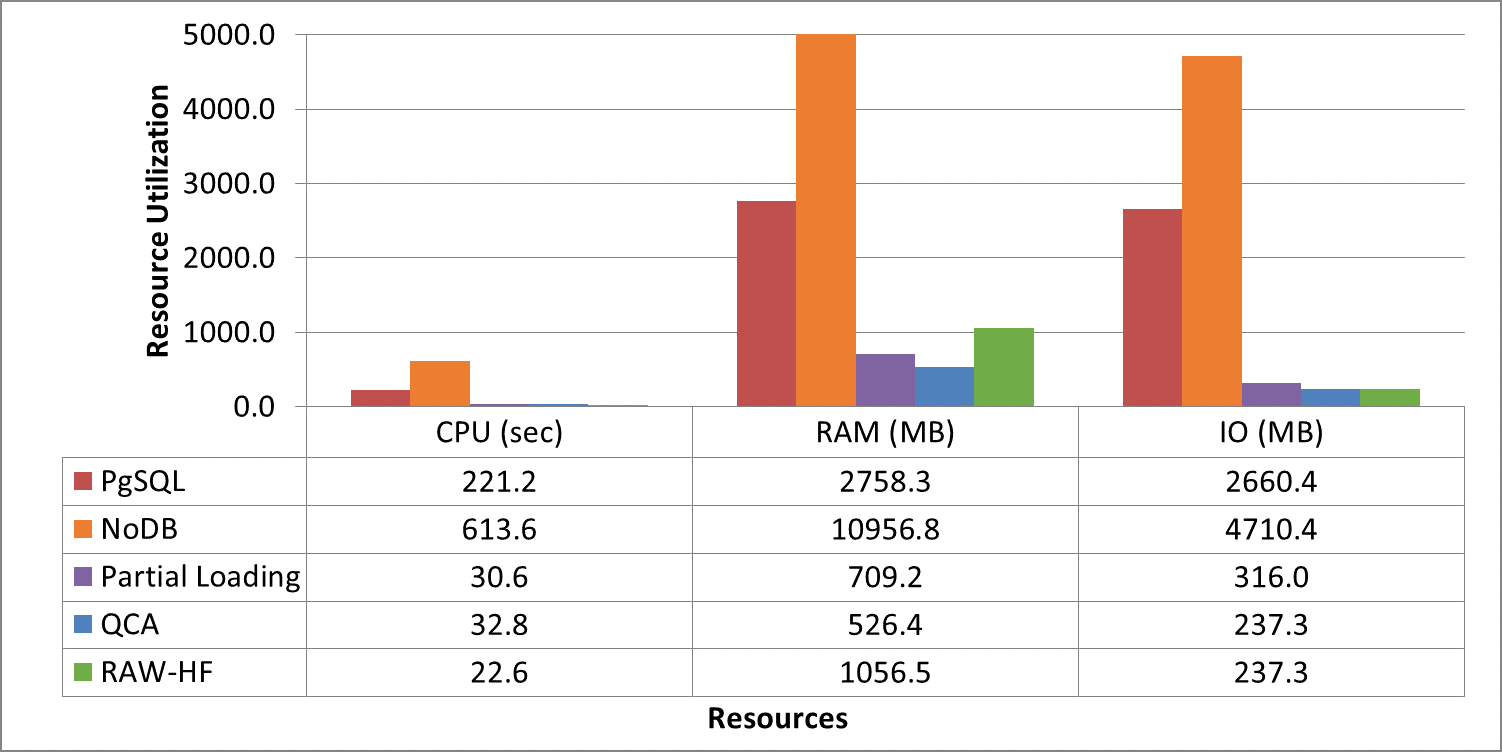}
    \caption{RAW-HF: Resource Utilization}
    \label{Fig_43_RAW_HF_Resource_Utilization}
\end{figure}

Figure \ref{Fig_43_RAW_HF_Resource_Utilization} shows the comparison of resources utilized by ORR, MUAR, and RAW-HF (ORR+MUAR) with NoDB  \cite{Alagiannis_Borovica_Branco_Idreos_Ailamaki_2012}, PgSQL DBMS  \cite{PostgreSQL}, and Partial loading technique  \cite{Zhao_Cheng_Rusu_2015b}. NoDB proposed a raw data query processing framework to process raw data in its place without loading. NoDB utilizes CPU for a longer time due to the high QET of CQs. RAM utilization is more than double the size of actual raw data. Here, the 1M records dataset used in experiment utilized 4.7GB of space on IO device.  PgSQL is the better choice for data processing as it reduced the CPU, RAM, and IO utilization by 72.5\%, 74.8\%, and 43.5\%, respectively.

MUAR results showed that CPU utilization time is reduced by only 6.34\% because most of the time is spent in data loading process compared to PgSQL. MUAR can only utilize other CPU cores to execute read queries in parallel. Figure \ref{Fig_43_RAW_HF_Resource_Utilization} confirms that CPU utilization is reduced by 77\% only during query processing tasks due to parallel processing. The IO utilization stays the same as MUAR experiments used the original 1M SDSS dataset having 509 attributes. The QCA with WSAC technique in ORR reduced the required DB partition size(IO) by 91.08\%, reducing WET, CPU, and RAM utilization by 85.9\%, 85.1\%, and 80.9\%. 

The Partial Loading technique \cite{Zhao_Cheng_Rusu_2015b} loaded only 10.6\% of original data into DBMS, which reduced the WET time by 88.1\% compared to NoDB. It also reduced the CPU and RAM utilization by 81.3\% and 86.4\%. The RAW-HF experiments combined ORR and MUAR techniques, which showed a 32.8\% increase in RAM utilization compared to the Partial Loading technique due to parallel processing of queries. However, RAW-HF improved DLT, QET, WET, CPU, and DB Size(IO) requirements by 29.5\%, 12.9\%, 26.14\%, 26.14\%, 24.92\% compared to Partial Loading technique \cite{Zhao_Cheng_Rusu_2015b} executing all read queries in parallel after data loading is complete. The maximum CPU utilization reached 94\% for RAW-HF while executing the given query workload. However, due to the 12 query workload, the average CPU utilization stayed at 31\% as most of the CPU time goes into data loading operations, which utilized a single CPU core. 

\subsection{RAW-HF for Different Datasets} \label{s_9_10_RAW_HF_Different_Datasets}

\begin{figure}[htbp]
    \centering
    \includegraphics[width=\textwidth]{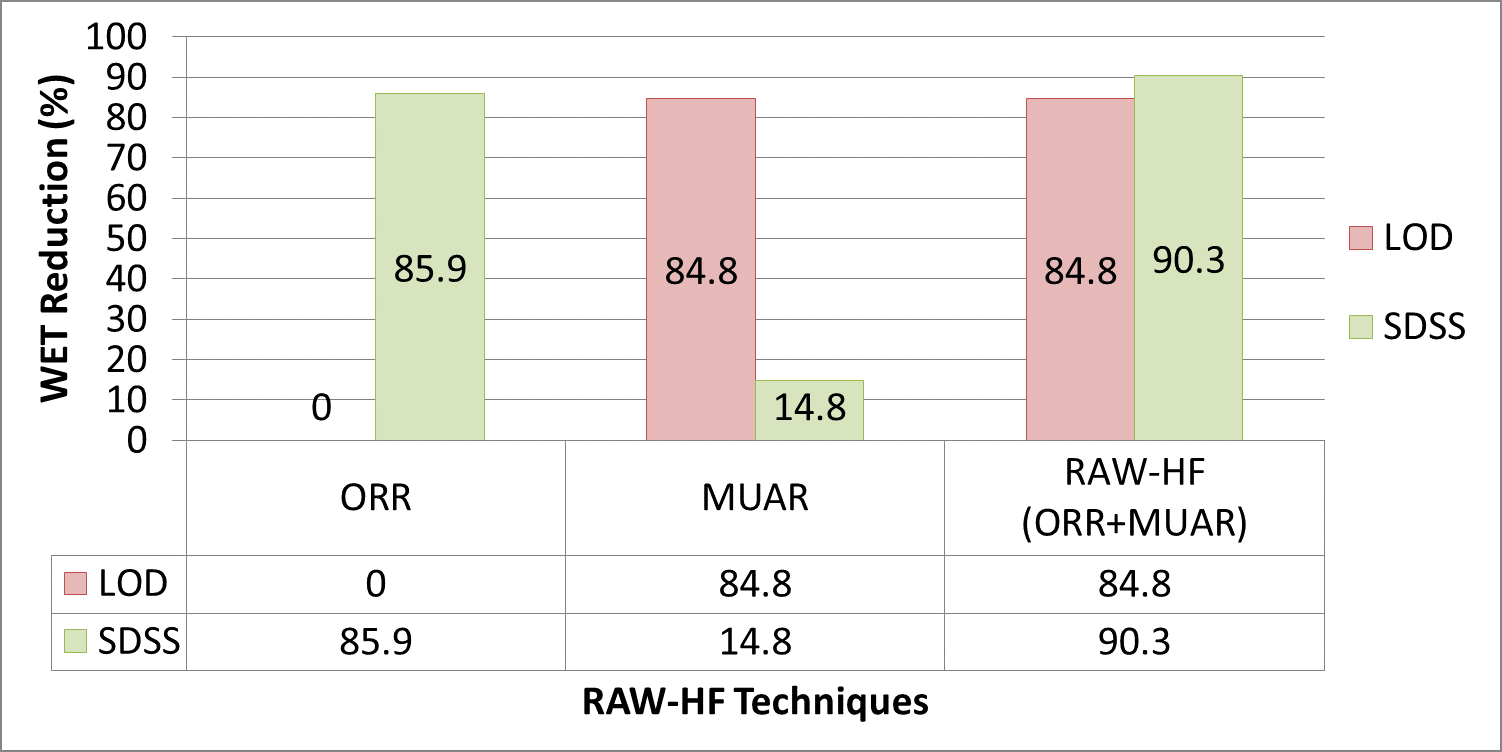}
    \caption{Impact of RAW-HF on WET for LOD \& SDSS datasets}
    \label{Fig45_RAWHF_for_Different_Datasets}
\end{figure}

 This section discusses the impact of RAW-HF on WET for different types of datasets like LOD \& SDSS. Table \ref{Table_18_ORR_FAA_FAL} compares LOD and SDSS datasets based on the Fraction of Attributes Accessed (FAA) by workload queries and the Fraction of Attributes Loaded (FAL) by RAW-HF. It can be seen that SDSS is a broad table dataset. All the SDSS workload queries access only 10.6\% of attributes. On the other hand, LOD dataset is a narrow table dataset containing only three attributes. Due to fewer attributes, almost all queries use two or more attributes. The impact of broad and narrow tables and queries accessing only small part or entire of the dataset can be seen in the ORR results for SDSS in Figure \ref{Fig45_RAWHF_for_Different_Datasets}. 

 \begin{table}[htbp]
\caption{ORR: Fraction of Attributes Accessed (FAA) and Loaded (FAL) }
\label{Table_18_ORR_FAA_FAL}

\begin{tabular}{|P{0.085\linewidth}|P{0.14\linewidth}|P{0.14\linewidth}|P{0.08\linewidth}|P{0.14\linewidth}|P{0.08\linewidth}|P{0.08\linewidth}|P{0.08\linewidth}|P{0.082\linewidth}|}
\hline
     & Total   Attributes & Accessed   Attributes & FAA   (\%) & Loaded   Attributes & FAL   (\%) & DLT   (\%) & QET   (\%) & WET   (\%) \\ \hline
\textbf{SDSS} & 509                & 54                    & 10.6       & 34                  & 6.7        & 90.9       & 87.9       & 85.9       \\ \hline
\textbf{LOD}  & 3                  & 3                     & 100.0      & 3                   & 100.0      & 0          & 0          & 0          \\ \hline
\end{tabular}
\end{table}

The ORR phase of RAW-HF uses vertical partitioning methods to reduce DLT and improve QET by accessing only required fractions by creating database and raw file partitions. RAW-HF only loads attributes required by complex queries to reduce DLT time, similar to QCA \cite{Patel_Bhise_2022}. This helps datasets like SDSS, which requires only a small fraction of the dataset (10.6\%) to answer queries by loading only 6.7\% of attributes. 
The remaining 93.3\% of attributes are not loaded into DBMS, reducing WET by 85.9\% for the SDSS dataset. On the other hand, vertical partitioning cannot help datasets like LOD that access 100\% of attributes. Therefore, the WET reduction achieved by applying ORR phase is 0\% for LOD. However, the MUAR achieves 84.8\% reduction in WET by efficiently utilizing existing CPU and RAM resources for complex queries. 
In summary, RAW-HF can be applied to  different types of real-world datasets to achieve significant reduction in WET by combining ORR and MUAR.



\section{Comparison with State-of-the-art} 
This section compares RAW-HF performance with state-of-the-art task scheduling, resource allocation, and partitioning techniques.

\subsection{RAW-HF for Complex Ad-hoc queries} \label{s_9_8_RAW_HF_CQ_Adhoc_Comparision}
\begin{figure}[htbp]
\centerline{\includegraphics[width=\textwidth]{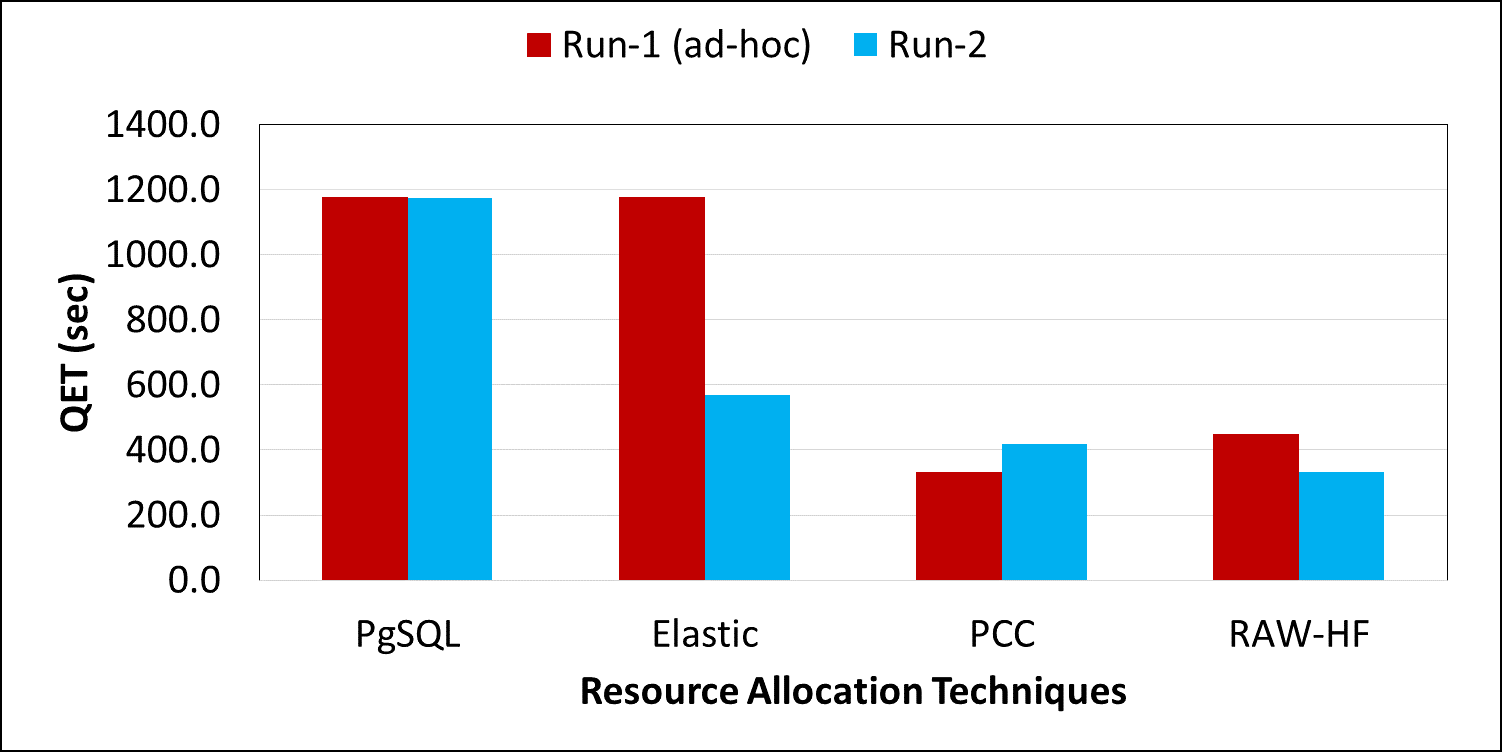}}
\caption{RAW-HF: Complex Query QET Comparison with state of the art }
\label{Fig_11_MUAR_CompareStateOfArt}
\end{figure}

This section discusses how RAW-HF handles complex ad-hoc queries. RAW-HF differentiates SQ and CQ queries in real-time and executes them using appropriate tools. The default resource allocation can not provide the lowest query response time, as discussed earlier in MUAR module.  MUAR module of RAW-HF tries to increase the utilization of existing system resources by allocating additional RAM resources to CQs to improve  QET \& overall WET. 
Figure \ref{Fig_11_MUAR_CompareStateOfArt} \& Table \ref{Table_16_RAW-HF_Technique_Comparison_with_State_of_the_art} 
 presents the comparison of RAW-HF with state-of-the-art dynamic or ML techniques based on parameters like QET, real-time resource utilization monitoring, whether the technique divides single query tasks for parallel processing, uses lightweight algorithms, and its ability to manage complex ad-hoc queries.

Figure \ref{Fig_11_MUAR_CompareStateOfArt} compares 1st and 2nd run QET of Q14 achieved by MUAR with PgSQL configured to allocate default resources, Elastic\cite{Raza_Chrysogelos_Anadiotis_Ailamaki_2020}, and  PCC\cite{Pimpley_Li_Sen_Srinivasan_Jindal_2022}. 
Experiments have been performed with multiple complex queries like Q10 \& Q14, which wrote 8-10GB of intermediate join results to disk.  
For the first run, PgSQL \& Elastic allocates default resources, i.e., 4MB \textit{work\_mem}. At the same time, MUAR allocates 1.8GB of work memory by analyzing query complexity and available RAM to improve 1st query run performance by 62\%. At the same time, PCC may allocate 8GB-10GB of RAM resources to achieve best performance during 1st query run as it over-allocates resources during initial runs in the serverless cloud. During 2nd run, Elastic resource allocation QET results are achieved by allocating only 500MB of work memory enough to reduce OLAP (complex query) QET by 50\%. 
PCC uses past data to train ML models with multiple features to allocate optimal (3.4GB) during the 2nd run with 12-20\% estimation error.  During 2nd query run executed on the same 7M records dataset, MUAR allocated 10.6GB of work memory by adding previous work memory of 1.8GB \& 8.8GB of disk writes recorded during 1st run. MUAR uses a simple linear equation that considers fewer parameters like join count, dataset size, and past disk writes to achieve the best performance with 15-20\% estimation error. This makes MUAR lightweight and faster compared to ML techniques.   
 The main memory utilization of MUAR is 3x to 20x higher than PCC and Elastic. In summary, MUAR can find the best resource allocation value for work memory parameter, which helps in achieving the lowest QET for a given query with single past query run data as PCC\cite{Pimpley_Li_Sen_Srinivasan_Jindal_2022}.

\subsection{RAW-HF Comparison with State-of-the-art} \label{s_9_9_RAW_HF_Comparision}

 Performance of each individual RAW-HF module has been presented in earlier works \cite{Patel_Bhise_2019,Patel_Bhise_2023a,Patel_Yadav_Bhise_2022,Patel_Bhise_2022,MUAR}. This section presents a comparison of the RAW-HF technique with other state-of-the-art techniques. The RAW-HF techniques compared with NoDB \cite{Alagiannis_Borovica_Branco_Idreos_Ailamaki_2012}, 
 Slalom \cite{Olma_Karpathiotakis_Alagiannis_Athanassoulis_Ailamaki_2020}, 
 DBMS \cite{PostgreSQL}, Partial Loading \cite{Zhao_Cheng_Rusu_2015b}, 
  PDC \cite{Tang_Byna_Dong_Koziol_2020}, and PCC \cite{Pimpley_Li_Sen_Srinivasan_Jindal_2022}. 
 NoDB is an open source in-situ processing engine with main memory caching and indexing features \cite{Alagiannis_Borovica_Branco_Idreos_Ailamaki_2012}. Although Slalom is an improvement over NoDB, it is not available as an open source tool  \cite{Olma_Karpathiotakis_Alagiannis_Athanassoulis_Ailamaki_2020}. PostgreSQL (PgSQL) is a widely used open source DBMS  \cite{PostgreSQL}. The Partial Loading technique proposes distributing dataset partitions among DBMS, and raw format considering storage resource limitations \cite{Zhao_Cheng_Rusu_2015b}. 
  PDC proposes to cache the dataset partition summaries and distribute query tasks to relevant nodes \cite{Tang_Byna_Dong_Koziol_2020}. PCC proposes using a performance characteristic curve to allocate appropriate resources to frequent queries  \cite{Pimpley_Li_Sen_Srinivasan_Jindal_2022}.  

\begin{table}[htbp]
\caption{RAW-HF Technique Comparison with State-of-the-art}
\label{Table_16_RAW-HF_Technique_Comparison_with_State_of_the_art}
\begin{tabular}{|P{0.025\linewidth }|P{0.18\linewidth }|P{0.08\linewidth }|P{0.08\linewidth }|P{0.08\linewidth }|P{0.09\linewidth }|P{0.07\linewidth }|P{0.08\linewidth }|P{0.24\linewidth }|}
\hline
\textbf{\#}                       & \textbf{Technique / Tool}                       & \textbf{Partiti-oning}              & \textbf{DBMS Data \%}               & \textbf{Work-load Aware}            & \textbf{Ad-hoc queries}             & \textbf{RUA} & \textbf{Multi-format Join}          & \textbf{Remarks}                                                                              \\ \hline
1                                 & PostgreSQL (PgSQL)  \cite{PostgreSQL}                     & -                                  & 100                                 & No                                  & -                                   & No                                  & No                                  & High DLT.  Low QET \& Resource Utilization.                                                   \\ \hline
2                                & NoDB (PostgresRaw)   \cite{Alagiannis_Borovica_Branco_Idreos_Ailamaki_2012}                    & -                                  & 0                                   & No                                  & NA                                  & No                                  & No                                  & Required More Memory. High QET.                                                               \\ \hline
3                                 & Slalom \cite{Olma_Karpathiotakis_Alagiannis_Athanassoulis_Ailamaki_2020}                                & Logical HP                         & 0                                   & Yes                                 & Yes                                 & No                                  & No                                  & Requires less Memory. Adapts to workload changes.                                             \\ \hline

4                                 & Partial Loading \cite{Zhao_Cheng_Rusu_2015b}              & VP                                 & 10.6                                & Yes                                 & -                                   & Yes                                 & No                                  & Technique Not Lightweight                                                                     \\ \hline
5                                 & PDC \cite{Tang_Byna_Dong_Koziol_2020} (ODBMS)                            & HP                                 & 0                                   & Yes                                 & No                                  & No                                  & No                                  & Resources Underutilized, Distributed System                                                   \\ \hline
6                                 & PCC (Cloud/ Serverless) \cite{Pimpley_Li_Sen_Srinivasan_Jindal_2022}                 & -                                  & 100                                 & Yes                                 & No                                  & Yes                                 & No                                  & Resource Intensive, Distributed System                                                        \\ \hline
{\color{blue} \textbf{{7}}} & {\color{blue} \textbf{RAW-HF (Hybrid)}} & {\color{blue} \textbf{VP}} & {\color{blue} \textbf{6.7}} & {\color{blue}\textbf{Yes}} & {\color{blue} \textbf{Yes}} & {\color{blue} \textbf{Yes}} & {\color{blue} \textbf{Yes}} & {\color{blue} \textbf{Lightweight Technique, Can be extended to a distributed setup}} \\ \hline
\end{tabular}
\end{table}

\begin{table}[]
\caption{RAW-HF Performance Parameters Comparison (SDSS)}
\label{Table_17_RAW_HF_Performance_Parameters_Comparison}
\centering
\begin{tabular}{|P{0.025\linewidth }|P{0.15\linewidth }|P{0.12\linewidth }|P{0.12\linewidth }|P{0.12\linewidth }|P{0.12\linewidth }|P{0.14\linewidth }|P{0.12\linewidth }|P{0cm}}
\hline
\multicolumn{1}{|c|}{\textbf{\#}} & {\textbf{Technique/ Tool}}               & \multicolumn{3}{|c|}{ \textbf{Query Performance \%}}                                                                            & \multicolumn{3}{|c|}{ \textbf{Resource Utilization \%}}                                                                                            \\ \cline{3-8} 
                                  & {\color{blue} }                                     & {{DLT (sec)}}                            & {QET (sec)}                           & {WET (sec)}       & {CPU (sec)}                            & {RAM (MB)}                               & {IO (MB)}          \\ \hline
1                                 & PgSQL \cite{PostgreSQL}                                             & {\textbf{188.63  \color{blue}(90.88\%)}}           & {\textbf{44.7 \color{blue}(87.92\%)}}            & \textbf{233.33 \color{blue}(90.31\%)}            & {\textbf{233.33 \color{blue}(89.78\%)}}           & {\textbf{2758.3  \color{blue}(61.70\%)}}            & \textbf{2660.4 \color{blue}(91.08\%)}            \\ \hline
2                                 & NoDB \cite{Alagiannis_Borovica_Branco_Idreos_Ailamaki_2012}                                               & {0}                                    & {\textbf{613.59 \color{blue}(99.12\%)}}          & \textbf{613.59 \color{blue}(96.32\%)}           & {\textbf{613.59 \color{blue}(96.32\%)}}           & {\textbf{10956.8 \color{blue}(90.36\%)}}            & \textbf{4710.4 \color{blue}(94.96\%)}            \\ \hline

3                                 & Partial Loading \cite{Zhao_Cheng_Rusu_2015b}                                    & {\textbf{24.4 \color{blue}(29.51\%)}}             & {\textbf{6.2 \color{blue}(12.90\%)}}             & \textbf{30.6 \color{blue}(26.14\%)}             & {\textbf{30.6 \color{blue}(26.14\%)}}             & {\textbf{709.2 \color{red}(+32.87\%)}}              & \textbf{316.0 \color{blue}(24.92\%)}             \\ \hline
{\color{blue} \textbf{4}} & {{\color{blue} \textbf{RAW-HF}}} & {{\color{blue} \textbf{17.2}}} & {{\color{blue} \textbf{5.4}}} & {{\color{blue} \textbf{22.6}}} & {{\color{blue} \textbf{22.6}}} & {{\color{blue} \textbf{1056.5}}} & {\color{blue} \textbf{237.3}} \\ \hline
\end{tabular}
\end{table}

 Table \ref{Table_16_RAW-HF_Technique_Comparison_with_State_of_the_art} presents a comparison of state-of-the-art raw data query processing techniques with RAW-HF. DBMS Data\% shows the percentage of original dataset loaded into DBMS by the technique or data processing tool. Resource Utilization Aware (RUA) shows whether the technique considered resource utilization information or not. The Multi-Format Join column represents whether the tool can execute join queries on data residing in raw and database formats. NoDB eliminates DLT by querying raw files. However, QET time and main memory utilization are very high. Slalom is an improvement over NoDB. It logically partitions raw files and adapts to workload changes using less main memory than NoDB. The PgSQL reduces the QET time at the cost of high DLT. NoDB, PgSQL, and PCC do not use partitioning techniques and require the entire dataset in a single format. Partial loading and RAW-HF use hybrid systems, so both can query multi-format data. The SCANRAW tool used to implement Partial Loading techniques can not join data existing in database and raw partition. Therefore, the multi-format (MF) join feature is not present. Whereas RAW-HF uses NoDB as an extension to PgSQL (PostgreSQL), allowing execution of join queries on the database and raw format. Therefore, the multi-format join feature is present in RAW-HF.

 Partial loading, PDC, PCC, and RAW-HF are workload aware. PCC uses workload information to identify appropriate resources, while others use workload information to partition the dataset. Only RAW-HF supports allocating appropriate resources to ad-hoc queries based on query complexity. While NoDB, Slalom, PgSQL, and Partial loading techniques allocate static resources to all the queries, including the ad-hoc ones. PDC uses a static partition of main memory (50\%) to keep lookup tables. PCC needs historical data for allocating appropriate resources to each query. However, it does not work well for ad-hoc queries. In comparison, RAW-HF performs workload-aware partitioning, and resource allocation is done based on query complexity. Therefore, RAW-HF is capable of allocating appropriate resources to ad-hoc queries. 

 The Partial Loading technique \cite{Zhao_Cheng_Rusu_2015b}, and RAW-HF partition the raw dataset into a database and raw partitions considering memory or storage budget. The storage budget limits the amount of data that needs to be loaded into DBMS. The Partial Loading technique tries to load attributes that cover maximum number of workload queries. RAW-HF uses lightweight ORR and MUAR modules to partition, schedule tasks, and allocate resources to reduce WET. It is not lightweight, as it requires attribute access time from database \& raw formats, data loading time, workload analysis, and other values to find the optimal partitions for hybrid systems, increasing algorithm execution time (AET). 
 The PCC uses historical resource utilization of queries. RAW-HF proposed to load only attributes required by complex queries to reduce DLT time and required to load only 6.7\% of data. The multi-format join feature present in RAW-HF can help in achieving 0\% replication for a single node when required. This means ad-hoc queries can access data loaded into DBMS and raw data to answer a query. In comparison, Partial Loading may load 10.6\% of the dataset accessed by workload queries when enough storage budget is available. PCC and PDC are implemented for cloud based systems making them distributed systems. Although RAW-HF is implanted on a single machine, it can be used for cloud based systems or extended to distributed setup.

 Figure \ref{Fig_42_WET_Comparison} \& \ref{Fig_43_RAW_HF_Resource_Utilization} presented comparison of these techniques with RAW-HF for SDSS dataset. Table \ref{Table_17_RAW_HF_Performance_Parameters_Comparison} shows the RAW-HF performance parameters comparison with state-of-the-art techniques. The table shows the time and resources required by techniques to execute the given workload on the SDSS dataset of 1M records. NoDB accessed the actual raw dataset file of size 4.7GB. PgSQL loaded the entire dataset into DBMS, reducing disk space by 43.5\%. Partial Loading technique \cite{Zhao_Cheng_Rusu_2015b} loaded only the data required by workload queries. The improvement achieved by RAW-HF in each performance parameter is written below the actual data in parentheses in blue color. At the same time, a decrease in performance is shown in red. It can be seen that RAW-HF improved WET by 26.14 to 96.32\% compared to others. The resource utilization RAW-HF reduced CPU and Disk space utilization for loaded data by 26.14\% and 24.92\%. But, RAM utilization is increased by 32.87\% compared to the Partial Loading technique.

\section{Conclusion}
The RAW-HF framework presented in this paper addresses the challenges of resource availability and workload optimization in hybrid systems. By combining lightweight partitioning, task scheduling, and query-specific resource allocation, RAW-HF aims to improve the efficiency of hybrid systems, reducing data-to-first-query time, DLT, QET, and WET.
RAW-HF performance has been demonstrated using scientific experiment datasets like SDSS and LOD. RAW-HF being resource-efficient helps process queries on large datasets faster. 

A comparison of the RAW-HF technique and performance with state-of-the-art techniques is presented. RAW-HF allows the execution of join queries on data stored in DBMS and raw format. At the same time, the Partial loading technique does not support the execution of join queries on data residing in multiple formats. RAW-HF never loads partitions used by simple queries, thereby reducing data loading requirements in ORR phase by 85.9\% for SDSS dataset. 
Unlike PCC, MUAR algorithm in MUER module allocates appropriate resources to ad-hoc queries by avoiding time-consuming offline analysis. It reduced WET by 84.8\% for LOD dataset. 
The RAW-HF reduced the total workload execution time by 26\% and 96\% compared to the state-of-the-art Partial Loading and NoDB techniques. The overall CPU, RAM, and IO resource utilization has been improved by 61-91\% over PostgreSQL DBMS. Partial loading technique requires 33\% lesser RAM than RAW-HF, but it needs 24\% more IO to achieve its best performance. Results analysis has shown that ORR phase works better for broad table datasets like SDSS, while MUAR is capable of improving WET for workloads with complex multi-join queries like LOD. RAW-HF reduced WET by 84.8\% for SDSS and 90.3\% for LOD datasets compared to traditional DBMS system  PostgreSQL. 

\end{document}